\documentclass[journal=ancac3,manuscript=article,layout=twocolumn]{achemso}

\usepackage[version=3]{mhchem} 
\usepackage[]{amsmath} %
\usepackage[]{color} 

\author{Eva G. Noya}
\affiliation[CSIC]{Instituto de Qu\'{i}mica F\'{i}sica Blas Cabrera, Consejo Superior de Investigaciones Cient\'{i}ficas, CSIC, Calle Serrano 119, 28006 Madrid, Spain}
\email{eva.noya@iqf.csic.es}
\author{Jonathan P.~K. Doye}
\affiliation{Physical and Theoretical Chemistry Laboratory, Department of Chemistry, University of Oxford, South Parks Road, Oxford, OX1 3QZ, United Kingdom}
\email{jonathan.doye@chem.ox.ac.uk}

\title[An \textsf{achemso} demo]
  {A one-component patchy-particle icosahedral quasicrystal}

\begin{document}







\begin{abstract}

Designing particles that are able to form icosahedral quasicrystals (IQCs) and
that are as simple as possible is not only of fundamental interest but is also
important to the potential realization of IQCs in materials other than metallic
alloys. Here we introduce one-component patchy-particle systems that in simulations 
are able to form face-centred IQCs that are made up of interconnected icosahedra. 
The directional bonding of the particles facilitates the formation of a network
of bonds with icosahedral orientational order and hence quasiperiodic positional
order. 
The assembled quasicrystals have similar energies to periodic approximants but
are entropically stabilized by phason disorder. Their long-range quasiperiodic 
order is confirmed by a higher-dimensional analysis. These materials, which are predicted to have an almost spherical photonic band gap, can potentially be realized via protein design and DNA origami particles.
\end{abstract}

\section{Introduction}

Quasicrystals are structures with long-range structural order (as evidenced by sharp peaks in a diffraction pattern) but with no periodically repeating unit. This lack of translational symmetry allows quasicrystals to exhibit symmetries not possible in periodic crystals. For example, the first quasicrystal discovered in a material, namely an alloy of aluminium and manganese, had icosahedral symmetry \cite{Shechtman84}. 
Subsequently, many other examples of icosahedral quasicrystals were discovered in metallic alloys \cite{Janot94,Takakura07}, but intriguingly never in any other type of material.

Whether icosahedral quasicrystals can be realized in other materials is thus an open question. One way to begin to address this question is to use simulations and theory to better understand the requirements for particles to assemble into quasicrystals. One feature of quasicrystals is that indexing their diffraction patterns requires (at least) two integers per quasicrystalline dimension and in icosahedral quasicrystals the two inverse length scales are related by the golden ratio $\tau=(1+\sqrt{5})/2$. Thus, one successful approach to design model particles that are capable of assembling into IQCs is to use isotropic potentials with complex radial forms that have features at multiple length scales \cite{Subramanian16,Ratliff19,Engel15,Damasceno17,Dshemuchadse21,Pan23}. 
However, how to realize particles with these complex potentials is less clear.

An alternative design strategy is to instead use directional interactions that favour the formation of a phase with the desired global symmetry. This approach was first used to produce ``patchy'' particles that can assemble into dodecagonal quasicrystals \cite{vanderLinden12,Reinhardt16,Tracey21} and more recently was extended to produce
patchy-particle IQCs \cite{Noya21}.  These original designs often required torsional terms, but new design strategies have recently been proposed to avoid this requirement, which may be hard to realize in experimental systems, but at the cost of introducing further particles types\cite{Pinto24}. One potential advantage of patchy particles over isotropic models is that the methods developed to design DNA origami particles \cite{Tian20,Liu24,Posnjak24} or proteins \cite{Li23} that can form a variety of crystals through their directional interactions might be extendable to realize DNA or protein quasicrystals. Finding model IQC-forming systems that are as simple as possible, as well as being of fundamental interest, is also likely to aid their experimental realization.
In Ref.\ \citenum{Noya21} two types of patchy-particle IQC-forming system were developed, both of which were shown to be simplifiable to binary systems. Here, we go one step further by introducing a third patchy-particle system that can form an IQC. Remarkably only one particle type is required. 

\section{Results}

\subsection{Particle design}
Our basic design approach is to choose the patch geometry of the particles so that it matches the directions of the bonds in the different local environments in our target structure. For crystals this is relatively simple \cite{Tracey19}, however, there are two additional complexities for QCs: firstly, how to generate the target QC; secondly, as the number of different local environments in the IQC can be large, how to avoid the use of an unreasonably large number of particle types.

\begin{figure*}[t]
\centering
\includegraphics[width=0.99\linewidth]{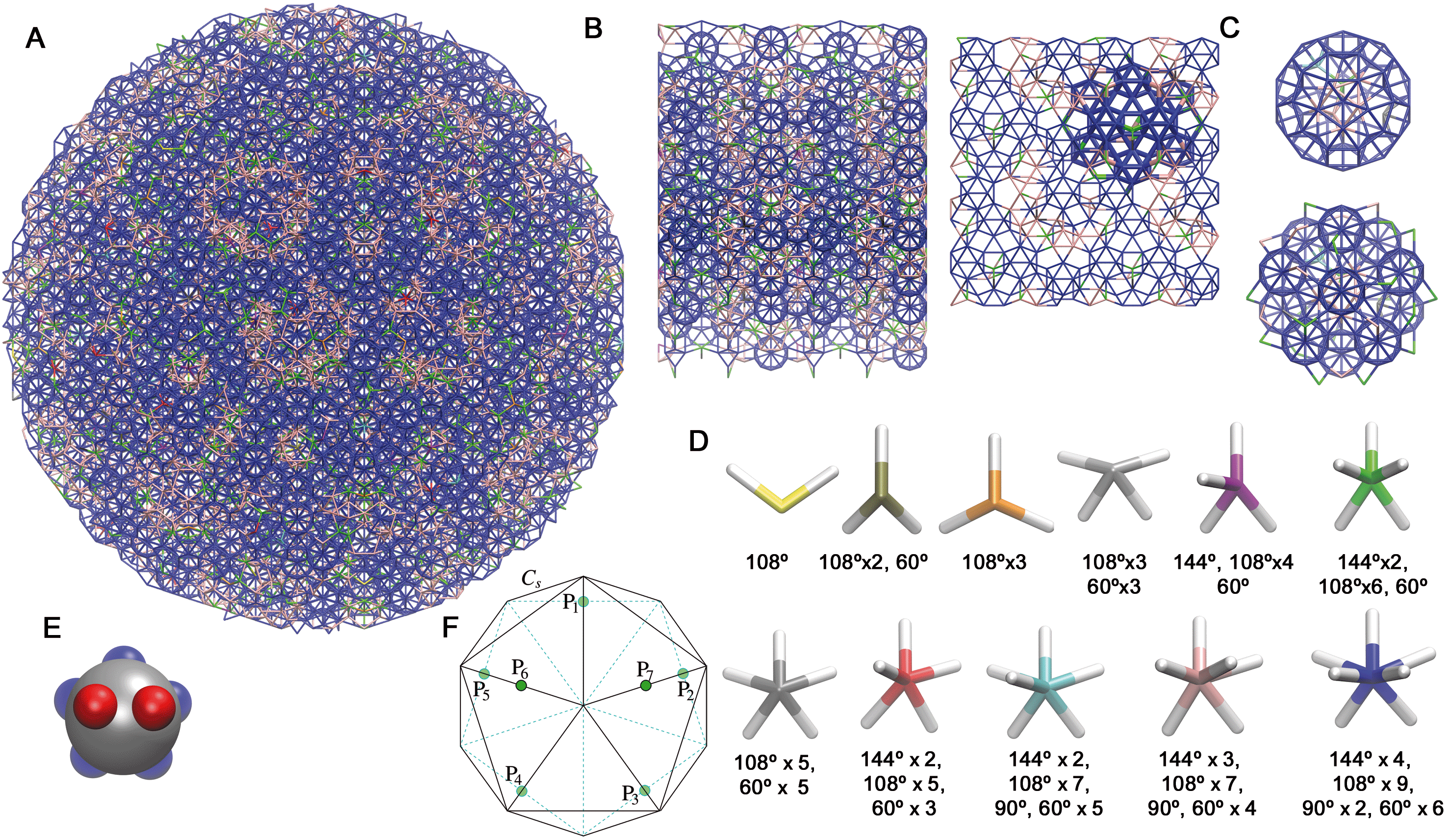}
\caption{From ideal quasicrystal to patchy-particle design. 
($A$) An ideal face-centred icosahedral/FCI quasicrystal viewed along a five-fold axis. 
($B$) A 3/2 rational approximant viewed along a pseudo-fivefold axis and a twofold axis. 
($C$) Example icosahedral clusters in the ideal quasicrystal: a rhombicosidodecahedron (top) and an icosahedra of icosahedra (bottom). These clusters are also present in the 3/2 approximant. The latter is also highlighted in $B$ with thicker bonds. 
($D$) Local environments in the ideal quasicrystal (with two or more neighbours). The colours match those used in $A$ and $B$. All environments can be considered as subsets of the 7-coordinate  environment. 
($E$) The patchy particle design that matches the 7-coordinate environment. 
($F$) The relationship between the symmetry of the patchy particle and the $I_h$ point group. The particle can be oriented so that its patch vectors $\mathbf{P_i}$ point along seven of the two-fold rotational axes of $I_h$.
Edges on the back faces of the icosahedra are dashed and cyan. Similarly, for patch vectors on the back faces, the colour shade is lighter and ringed in cyan rather than black.
}
\label{fig:FCI_ideal}
\end{figure*}

To generate an ideal target IQC we use the cut-and-project method \cite{Steurer09}. This involves the projection of a subset of the lattice points in a 6-dimensional hypercubic lattice onto an appropriately chosen 3-dimensional hyperplane. 
This hyperplane is termed the ``parallel'' space, and the 3-dimensional space orthogonal to this hyperplane is termed the ``perpendicular'' space.
In particular, lattice points are projected if a 3-dimensional volume in perpendicular space (termed the ``occupation domain'') centred on that point intersects the hyperplane. As parallel space represents an irrational cut 
through the hyperspace, the resulting structure cannot be periodic. 
This approach can be modified to generate a periodic crystal by instead choosing
a hyperplane that has a rational slope. These periodic crystals are termed rational approximants.
Previously, we have used body-centred and primitive hypercubic lattices \cite{Noya21}. Here, we instead use a face-centred lattice. 

An example ideal IQC produced by this approach is shown in Fig.\ \ref{fig:FCI_ideal}$A$. Also depicted is the 3/2 rational approximant to the IQC (Fig.\ \ref{fig:FCI_ideal}$B$). This approximant has 288 atoms per unit cell (285 disregarding zero-coordinated particles) and belongs to the $R3$ (146) space group.
A common local motif in both structures is the simple 12-particle icosahedron. 
In some places in the ideal IQC and the 3/2 approximant, these icosahedra are even arranged into an icosahedron of icosahedra (Fig.\ \ref{fig:FCI_ideal}$C$); however, more often the requirements of long-range order lead to clusters involving some defective or incomplete icosahedra. Consequently, the number of different environments in the ideal IQC is large. The eleven environments with a coordination number of 2 or more are illustrated in Fig.\ \ref{fig:FCI_ideal}$D$; the environment with a coordination number of 7 is most common (69\%) and 97\% of the environments have a coordination number that is greater than or equal to 5.
In the previous examples, we grouped together the environments that were subsets of each other and introduced particles corresponding to the highest coordination environment in each class \cite{Noya21}, the hope being that, even though not all the patches would be used in the IQC, it would still represent the preferred structure for the patchy-particle system.

Here, all eleven environments form one class all being subsets of the 7-coordinate environment. Therefore, we explored the self-assembly behaviour of one-component systems of the 7-patch particle depicted in Fig.\ \ref{fig:FCI_ideal}$E$ (designated the {7P} FCI model). Fig.\ \ref{fig:FCI_ideal}$F$ illustrates the relationship of the patch geometry to the icosahedral point group $I_h$. The particle can be oriented so that all 7-patches point along $C_2$ axes of the group. 
The patches form two sets. 
The first set $A$ consists of 5 patches related by one of the $C_5$ axes of $I_h$. These patches are typically involved in forming the intra-icosahedral bonds. In fact, particles just having these five patches have been extensively explored as models of particles able to form finite complexes with high symmetry; they are able to assemble from a monomeric gas into a gas of icosahedral clusters \cite{Wilber07,Wilber09,Wilber09b}.
The second set $B$ consists of two patches related by a mirror plane of $I_h$ and are mainly responsible for inter-icosahedral bonds; these form connections between the edges of the icosahedra. 

The interactions between particles are described by the same patchy-particle model as in Refs. \citenum{Tracey19,Noya21}. In
this pair potential, the interaction is described by a Lennard–Jones
repulsive core and an attractive tail modulated by angle and possibly torsion dependent functions. 
The angular modulation term
has a Gaussian form and is a measure of how directly two patches
point at each other. The standard deviation of the Gaussian $\sigma_\mathrm{ang}$
is a measure of the angular width of the patch.
The torsional modulation term describes the variation in the
potential as either of the particles is rotated about the interparticle
vector. It also has a Gaussian form that is centred at the preferred
torsional angle. The parameter $\sigma_\mathrm{tor}$ is a measure of the torsional
specificity of the interaction. To capture the symmetry of an environment, more than one preferred torsional angle can be defined.
Unless otherwise stated we use $\sigma_\mathrm{ang} = 0.3$ radians, as this provides a reasonable balance between the patches being sufficiently narrow to favour the target structure whilst not being so narrow that the kinetics of assembly is significantly hindered. When using
torsionally-specific interactions, we use $\sigma_\mathrm{tor} = 2\sigma_\mathrm{ang}$. However, we found that adding a torsional term was not necessary to achieve IQC assembly in the current systems.
 All patches interact with each other, but we use the relative
strength of the interactions between the two types of patches $\epsilon_{BB}/\epsilon_{AA}$ to help control the assembly. We
set $\epsilon_{AA} = 1$ and consider different values for $\epsilon_{BB}$. We always use
$\epsilon_{AB} = (\epsilon_{AA} + \epsilon_{BB})/2$. We use $\sigma_\mathrm{LJ}$
(the distance at which the Lennard–Jones potential is zero) as our
unit of length, and the Lennard–Jones well depth $\epsilon_\mathrm{LJ}$ as our unit
of energy.
Temperatures are given in reduced form,
$T^*= k_B T /\epsilon_\mathrm{LJ}$. 
The properties of the patchy interactions for the different models are fully tabulated in the SI, as well as 
the precise mathematical form of the potential for the ring patches. 

\begin{figure*}[t]
\centering
\includegraphics[width=.99\linewidth]{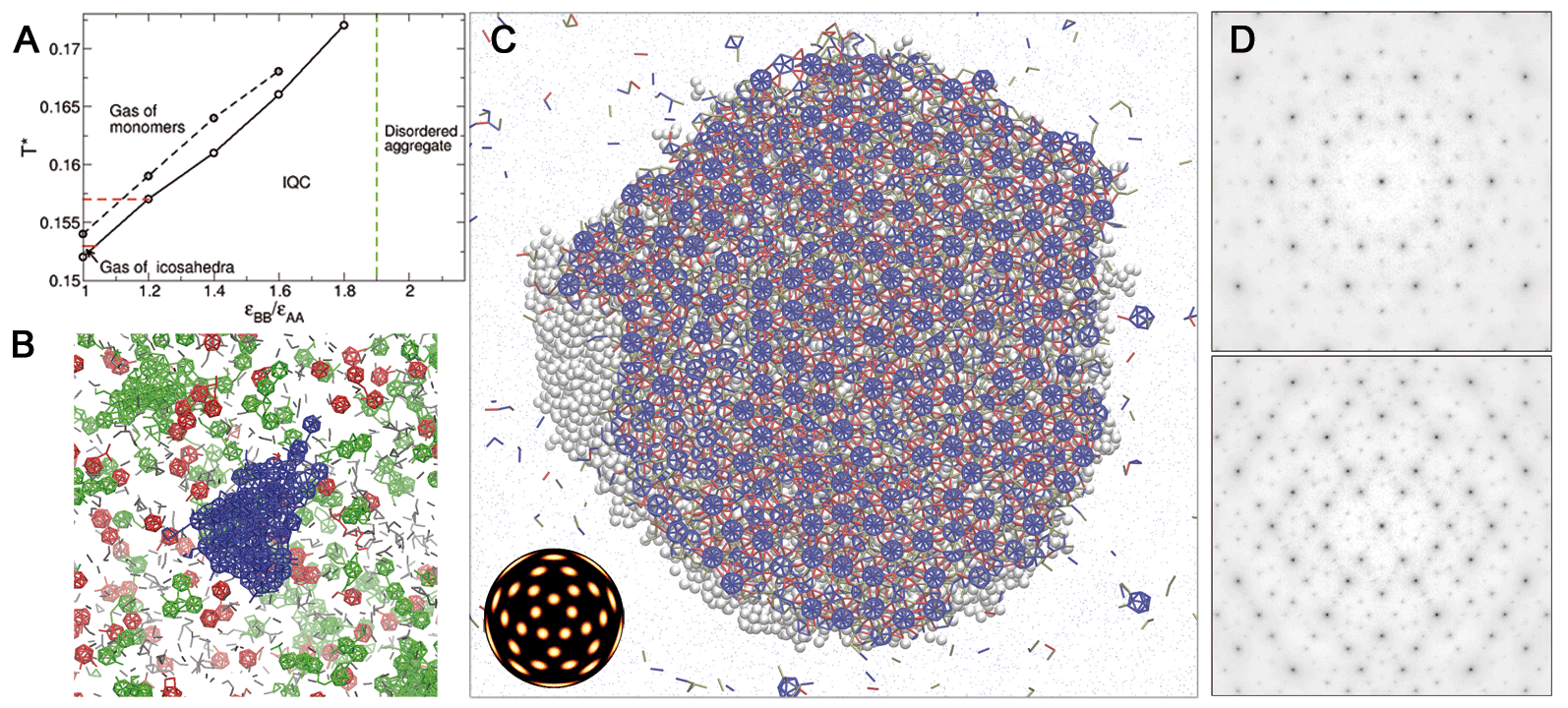}
\caption{Assembly of the {7P} FCI model. ($A$) Schematic assembly diagram as a function of $\epsilon_{BB}/\epsilon_{AA}$. Solid circles mark the highest temperature at which spontaneous assembly was observed for each $\epsilon_{BB}/\epsilon_{AA}$ on the time scale of our simulations, and open circles the temperatures at which the properties of the assembled IQCs were studied. The solid and dashed black lines are guides to the eyes. At $\epsilon_{BB}/\epsilon_{AA}$=1.0 the formation of the IQC was hindered by the assembly of multiple small aggregates. In this case we set the crystallization temperature as that for which the largest cluster was observed (that contained about 800 particles, shown in $B$).  For $\epsilon_{BB}/\epsilon_{AA} > 1.9$ particles assembled into a disordered aggregate. 
The solid red horizontal line demarcates the centre of the transition from a gas of monomers to a gas of icosahedra that is  estimated from simulations of particles with five patches of type $A$ (see Supplementary Material and Fig.\ S9 for further details). The dashed red line provides an indication of the width of this transition by marking the temperature at which 30\% of particles are in icosahedra. 
($B$) View of a typical configuration at $\epsilon_{BB}/\epsilon_{AA}=$1.0 and $T^*=0.152$. The largest cluster is shown with blue bonds, clusters containing more than 15 particles with green bonds, those with between 9 and 15 particles with red bonds and those with less than 9 particles with gray bonds. 
($C$) Cut through an assembled 45\,000-particle cluster obtained at
$\epsilon_{BB}/\epsilon_{AA} = 1.2$ and $T^*=0.159$. Bonds close to the cut plane are coloured using the scheme: $AA$ bonds in blue, $BB$ bonds in red, and $AB$ in tan. Other particles in the cluster are shown as grey spheres and particles in solution as grey dots. The BOOD shows that all the bonds are directed along the 2-fold axes of $I_h$. 
($D$) Diffraction pattern projected along the 5-fold and 2-fold symmetry axes.
}
\label{fig:FCI_assembled}
\end{figure*}

\subsection*{IQC self-assembly}

To explore whether our patchy-particle design could assemble into an IQC we used constant temperature Monte Carlo simulations starting from a low-density fluid. The temperature was chosen so that there would typically be nucleation and growth of a single cluster of the condensed phase. A key variable in the self-assembly behaviour is $\epsilon_{BB}/\epsilon_{AA}$ as this affects the relative favourability of intra- versus inter-icosahedral bonds. 

Fig.\ \ref{fig:FCI_assembled}$C$ shows a cut through a 45\,000 particle cluster grown at $\epsilon_{BB}/\epsilon_{AA}=1.2$. 
The diagram is coloured to highlight the icosahedra that are held together by $AA$ bonds. It can be seen that the icosahedra are all oriented with their local five-fold axes of symmetry out of the plane. The global orientational order is confirmed by the bond-orientational order diagram (BOOD). 
The BOOD shows that the bonds are oriented along the 2-fold directions of $I_h$, as expected from the patchy-particle geometry (Fig.\ \ref{fig:FCI_ideal}$F$). 
The quasicrystallinity of the cluster is confirmed by the diffraction patterns (Fig.\ \ref{fig:FCI_assembled}$D$); they both show that the cluster has long range order and that it possesses five-fold symmetry around the relevant axes. 
A comparison of the diffraction pattern to that for the ideal quasicrystal (Fig.\ S3) confirms that they share the same set of major peaks, and an indexing of the pattern (Fig.\ S4) confirms that the quasicrystal is face-centred icosahedral. 
Thus, we have discovered a one-component patchy-particle IQC-forming system. It is noteworthy that this was achieved without introducing a torsional component to the potential to disfavour possible  competing crystals; controlling just the geometry of the local coordination shell through the patch geometry is sufficient to ensure IQC formation. 
The lack of competing simple crystals may partly reflect just the number of patches; it has previously been noted that there is a paucity of crystals with an average coordination number close to 7 \cite{Dshemuchadse21}.

We obtained the most ordered IQC clusters at $\epsilon_{BB}/\epsilon_{AA}=1.2$. 
At $\epsilon_{BB}/\epsilon_{AA}=1.0$, however, IQC self-assembly was not observed. Under these conditions, instead of forming an IQC direct from the low-density fluid, the particles first start to assemble into isolated icosahedra. Although some of these icosahedra further assemble into smallish aggregates, the majority remain as icosahedra and the formation of an IQC is very difficult; Fig.\ \ref{fig:FCI_assembled}$B$ shows an example of the configurations that result. The problem is that when the icosahedra form the orientation of the $B$ patches on the surface will effectively be random and unlikely to match that needed for growth of the quasicrystalline phase. 
Although the system is not completely dynamically arrested --- e.g. there is a dynamic equilibrium between monomers and icosahedra --- the reorientational dynamics of the assembled particles is too slow for significant IQC formation on the simulation time scales. 
Note that this lack of assembly is purely due to kinetic factors; if the IQC grown 
at $\epsilon_{BB}/\epsilon_{AA}=1.2$ is then simulated at $\epsilon_{BB}/\epsilon_{AA}=1.0$ 
it is stable under the same conditions as in Fig.\ \ref{fig:FCI_assembled}$B$.

A schematic assembly diagram as a function of $\epsilon_{BB}/\epsilon_{AA}$ and temperature is shown in Fig.\ \ref{fig:FCI_assembled}$A$. It shows the regions of stability of the monomeric fluid, the fluid of icosahedra and the icosahedral quasicrystal, with the latter being increasingly stabilized with respect to the fluid phases as $\epsilon_{BB}/\epsilon_{AA}$ increases. Only beyond about $\epsilon_{BB}/\epsilon_{AA}=1.3$ is there a direct transition between the monomeric fluid and the quasicrystal. At $\epsilon_{BB}/\epsilon_{AA}=1.2$ there is still a competition between the formation of isolated icosahedra and the IQC, but unlike at $\epsilon_{BB}/\epsilon_{AA}=1.0$ it just slows down assembly rather than prevents it (Fig.\ S8$B$); indeed, the slower growth might be one of the reasons that the IQC at $\epsilon_{BB}/\epsilon_{AA}=1.2$ is relatively so ordered with its properties most closely matching that of a simulated ideal IQC (Table S4). IQC formation is observed up to $\epsilon_{BB}/\epsilon_{AA}=1.8$, but beyond this disordered aggregates lacking icosahedra instead form because of the comparative weakness of intra-icosahedral bonding.

\begin{figure*}
\centering
\includegraphics[width=0.97\linewidth]{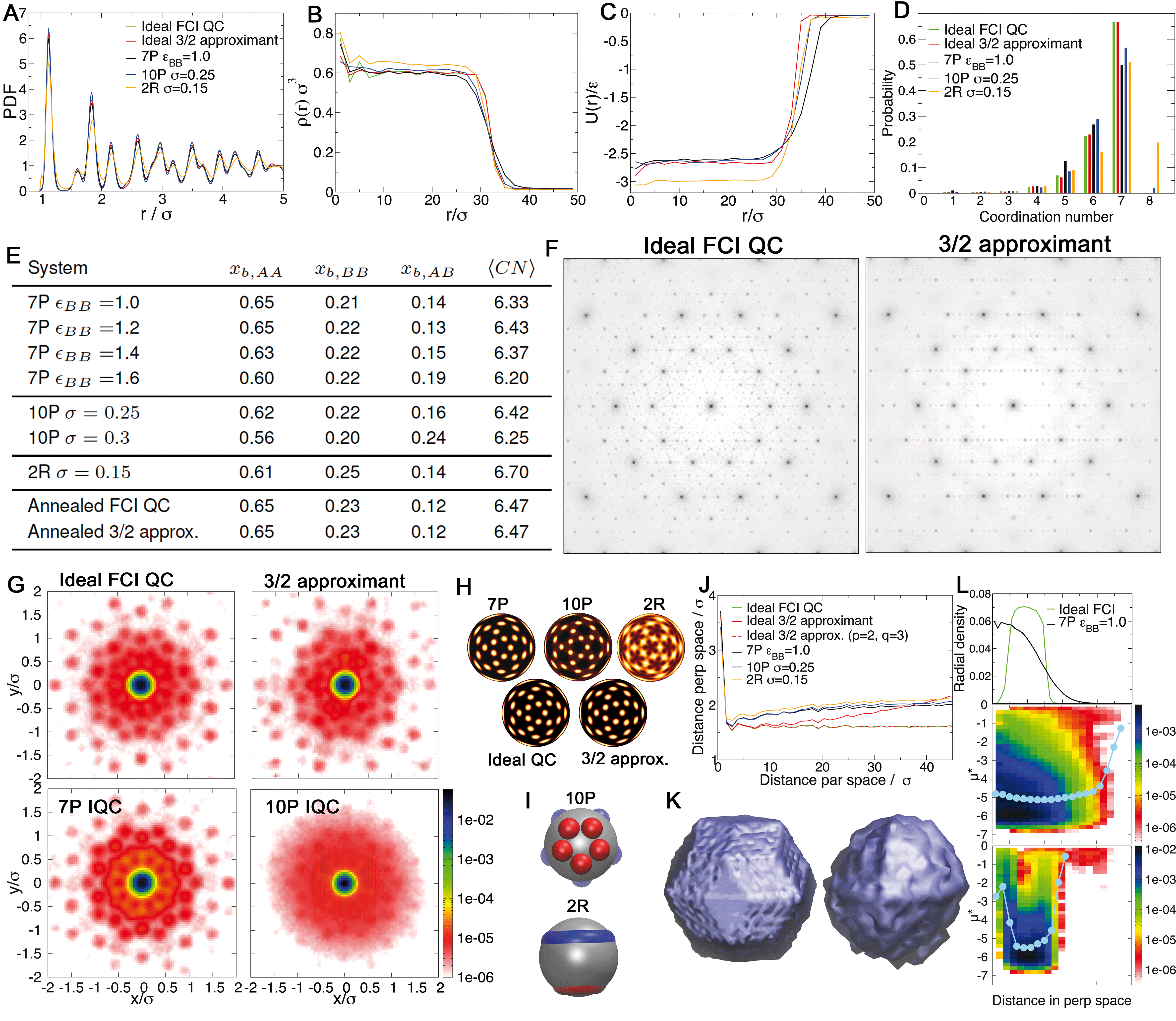}
\caption{Comparison of the assembled quasicrystals (7P IQC, 10P IQC and 2R IQC) and the ideal 
FCI QC and a 3/2 approximant simulated with the 7P model. In these five simulations, $\epsilon_{BB}/\epsilon_{AA}=1$ (the starting point for the FCI 7P annealing simulation was a cluster grown at $\epsilon_{BB}/\epsilon_{AA}=1.2$). 
($A$) Pair distribution functions. ($B$) Density $\rho(r)$ and ($C$) energy $U(r)$ as a function of the distance from the centre of the cluster $r$. ($D$) Coordination number distributions. ($E$) The average coordination number ($<CN>$) and the fraction of $AA$, $BB$ and $AB$  bonds ($x_{AA}$, $x_{BB}$ and $x_{AB}$). For the 7P model, results for interactions strengths $\epsilon_{BB}/\epsilon_{AA}=$1.0, 1.2, 1.4 and 1.6 (all with $\sigma_\mathrm{ang}=0.30$) are also given. For the 10P model data is shown for $\sigma_\mathrm{ang}=0.25$ and for $\sigma_\mathrm{ang}=0.30$ with a torsional term.  ($F$) Diffraction patterns of the annealed ideal FCI QC and the 3/2 rational approximant projected along the 5-fold axis. ($G$) Van Hove correlation functions. ($H$) BOODs for all 5 systems.
($I$) The 10P and 2R patchy particles.
($J$) Phason strain. For the approximant the distance in perpendicular space has zero slope if the lifting is done taking $p=$2 and $q=3$ in perp space. ($K$) Shape of the occupation domain in the annealed ideal FCI QC (left) and the assembled 7P IQC (right). ($L$) Radial density in perpendicular space (top panel). Probability density of the occupation domain as a function of distance in perpendicular space and particle's energy ($u^*$) for (middle panel) the 7P IQC ($\epsilon_{BB}/\epsilon_{AA}=1.0$) and for (bottom panel) the annealed ideal FCI QC. Blue line and symbols indicate the average energy as a function of distance in perp space.} 

\label{fig:properties}
\end{figure*}

In our search for the simplest IQC-forming patchy-particle system it would have been preferable if all the patch-patch interactions could be of equal strength. We therefore explored possible ways that the kinetic traps that prevent assembly at $\epsilon_{BB}/\epsilon_{AA}=1$ might be overcome. Our solution was to instead use a particle with five equivalent $B$ patches (Fig.\ \ref{fig:properties}$I$), named {10P} FCI model. Such particles still effectively have a maximum coordination number of seven as it is not possible for two adjacent $B$ patches to be simultaneously involved in strong bonds since the angle between the patches is $36^\circ$ and the neighbouring particles would be significantly displaced from the minimum energy positions to avoid overlaps (Fig.\ S10). 
However, now once an icosahedron forms the directions in which $BB$ bonds can form are no longer pre-determined, as any two of the five $B$ patches can be used. Our initial simulations with such particles led to the formation of a liquid droplet; note that it is well-established that increasing the number and width of patches increases the temperature of the liquid-vapour critical point \cite{Bianchi06}. We therefore made the patches narrower setting $\sigma_\mathrm{ang}=0.25$ radians. This led to the successful assembly of an IQC (Fig.\ S7). Although the addition of the extra 3 $B$ patches was motivated by their potential effect on the dynamics, they also affect the thermodynamics and assembly now actually occurs direct from the monomeric vapour (Fig.\ S8). 
This is because the increase in the number of $B$ patches entropically stabilizes the IQC (there are more possible particle orientations that are compatible with the IQC) but has no effect on the transition to a gas of icosahedra.
We also note that adding torsional interactions rather than making the patches narrower also sufficiently disfavours the formation of disordered aggregates that an IQC grows direct from a low-density fluid (Fig.\ S8).

Additionally, we tested the assembly of a third design in which each of the two sets of 5 patches in the {10P} model was replaced by a ring patch (Fig.\ \ref{fig:properties}$I$).  We designate this model {2R FCI}.
The width of the ring patches was reduced to $\sigma_\mathrm{ang}=0.15$ radians to avoid stabilization of the liquid phase. This less specific design is still able to form an FCI QC (Fig.\ S7), albeit somewhat less ordered than for the QCs assembled from the {7P} and {10P} models.

\subsection*{Analysis of annealed samples} 

Besides analysing the structural and dynamic properties of
the {7P}, {10P} and {2R} types 
of assembled IQCs, we also constructed patchy-particle versions of the ideal IQC and the 3/2 rational approximant and compared the properties of the five systems averaged over long annealing simulations (Fig.\ \ref{fig:properties}). The initial configurations were built by placing particles at the lattice sites of the ideal configurations of the IQC and the 3/2 rational approximant. Particle orientations were obtained by using a Monte Carlo code that, for each particle, searches for the orientation that minimizes the deviation of the particle's patches from the first coordination shell in the ideal configuration. These configurations are then linearly heated from a reduced temperature of 0.01 to 0.154 over a 1 million MC cycles simulation.  The ideal FCI QC and its rational approximant are then allowed to evolve for at least further 4 million MC cycles at $T^*$=0.154, using the 7P model with $\epsilon_{BB}/\epsilon_{AA}=$1. 
The structural properties of the assembled 7P, 10P, and the annealed ideal IQC and 3/2 approximant are very similar, including the radial distribution functions and the bulk densities and energies (Fig.\ \ref{fig:properties}$A-C$). 
The latter is particularly noteworthy, as it means that the IQC is likely to be thermodynamically stable over a significant range of temperature due to its greater entropy. The absence of a significant energetic advantage for the approximant also helps to further explain why it is never observed in the assembly simulations.

Consistent with the similar energies the average coordination number for the ideal IQC and the approximant
are effectively the same and only 8\% less than the maximum possible value of 7 (again helping to explain why no alternative crystal forms were observed). Furthermore, the most ordered examples of the assembled 7P and 10P IQCs have a coordination number that is only a bit less and with a similar proportions of inter- and intra-icosahedral bonding (Fig.\ \ref{fig:properties}$E$).
The 2R IQC differs from the other four systems in that the reduced angular constraints associated with the ring patches allow a significant fraction of particles to achieve a coordination number of 8, and hence a higher average coordination number, higher density and lower energy. The spots in the BOOD are also no longer isotropic, but have features associated with the angular vibrations about the particles' symmetry axes being energetically less costly (Fig.\ \ref{fig:properties}$H$).

The order in the approximant is very similar to that of IQCs. For example, the deviations in the BOOD from icosahedral symmetry are not apparent to the eye (in fact bonds are $\sim 5$\% more likely along the $C_2$ axes along $x$, $y$ and $z$ (Fig.\ S12)) and it is only in the weaker higher-order peaks in the diffraction pattern that the deviation from long-range five-order is evident (Fig.\ \ref{fig:properties}$F$).

The Van Hove correlation functions show that particle hops are possible in the IQCs and the approximant
and exhibit (approximate) icosahedral symmetry. 
It is also noticeable that the structure in the Van Hove correlation function for the 10P and 2R IQCs is significantly less well-defined; the extra patches in 10P and the ring patches in 2R seem to allow greater freedom in the relative particle motion (Fig.\ \ref{fig:properties}$G$). 
In the previous multi-component patchy-particle IQCs \cite{Noya21}, although the ``matrix'' particles exhibited similar mobility to the IQCs here, the particles associated with the (rhombic triacontahedral or dodecahedral) clusters tended to be relatively immobile, preventing significant structural change. The one-component nature of the current system is likely to allow easier structural relaxation; however, we should note that the particle mobility in the assembled quasicrystals decreases as a function of simulation time (Fig.\ S13), presumably due to the annealing out of mobility-facilitating defects.

Another way to analyse the assembled IQCs is to do the reverse of the cut-and-project procedure that was used to generate our starting ideal IQC. In this ``lifting'' procedure every particle in the IQC is mapped onto a lattice point in 6D space.  This can be achieved because the 6D interlattice vectors corresponding to each of the thirty bond directions (along the $C_2$ axes of $I_h$) in the ideal IQC are known. Thus, by iterating through the bond network all particles can be ``lifted''. Important to this mapping being well-defined for the assembled IQCs is the absence of dislocations in these systems. 

The lifted 6D coordinates occupy the face-centred lattice sites as expected. Indeed it follows that an IQC that exclusively has bonds along the two-fold directions must be face-centred, as a vector in parallel space that is along a two-fold direction can only be generated from a projected 6D lattice vector that has non-zero components along an even number of lattice directions.

This procedure allows the quasiperiodicity of our structures to be assessed. Fig.\ \ref{fig:properties}$J$ shows how the distance between the lifted lattice points in perpendicular space depends on their separation in parallel space. Quasicrystals are said to have zero phason strain if the slope of this plot is zero \cite{Engel15,Je21}.
This will trivially be the case for the ideal IQC generated by the cut-and-project method and still holds after this ideal IQC is annealed at finite temperature. 
By contrast, the 3/2 approximant has a clear linear phason strain, as expected from the different slope of the hyperplane used to generate this approximant. This phason strain is coherent in the sense that if the perpendicular distance was measured perpendicular to this rational hyperplane the slope would be then reduced to zero (the $p=2,q=3$ line in Fig.\ \ref{fig:properties}$J$).
For random tiling quasicrystals, the entropy is maximal for zero phason strain \cite{Henley91}. But if a quasicrystal has quenched-in disorder (e.g.\ due to rapid solidification) this inability to reach equilibrium can lead to a finite phason  strain \cite{Tsai08}. In our assembled IQCs the phason strain is approximately zero,  further confirming their quasicrystallinity (Table\ S5). 

The limiting values of the separation in perpendicular space are greater for the assembled IQCs than the ideal IQCs. This is as expected and reflects the greater phason disorder in the assembled systems (Table S5). The kinetically, and probably also thermodynamically, preferred state of the assemblies involves inherent disorder.

The occupation domain for the IQCs can be obtained by projecting the lifted lattice points into perpendicular space. For the ideal IQC the occupation domain retains its rhombic-triacontahedral shape on annealing (Fig.\ \ref{fig:properties}$K$) but for the assembled IQCs the occupation domain is larger and more diffuse consistent with its greater phason disorder. In Fig.\ \ref{fig:properties}$L$ we show how the energy of the particles depends on their distance from their centre of the occupation domain. Interestingly, for the assembled quasicrystals a particle's energy is relatively independent of its position within the occupation domain. This also occurs in the assembled IQCs from our previous work\cite{Noya21} (see Fig.\ S16), but contrasts with the IQCs in Ref.\ \citenum{Engel15}. 
This feature helps to explain why the assembled FCI QCs have a similar energy to the approximant and the ideal IQC even though they have significant phason disorder.
It also suggests that the driving force for their zero phason strain is likely to be entropic. This is somewhat similar to the random tiling model of quasicrystals \cite{Henley91}. However, the configurations of the IQCs cannot be simply divided into a set of tiles with identical particle decorations; instead the contributors to the configurational entropy are more diverse.

\section*{Conclusions}

In summary, we have demonstrated that it is possible to assemble IQCs from one-component patchy colloidal systems. In some sense,  our particle designs are somewhat akin to a patchy-particle ``einstein'' (an einstein is a monotile that can cover the plane aperiodically \cite{Smith23,Smith23b}). 
However, in our particle-based models, the assembled IQCs cannot be straightforwardly interpreted in terms of a few tiles with identical particle decorations that completely fill space. 

In the current FCI examples, the predominant motif is the 12-particle icosahedron. Thus, the assembled IQCs share some similarities to icosahedral glass models \cite{Stephens86,Stephens89} which are obtained by irreversible aggregation of icosahedra with fixed orientation. However, in the patchy-particle IQCs, there are also incomplete icosahedra and a small fraction of particles that are not associated with icosahedra. Furthermore, as the growth is not irreversible, the particles, and hence the icosahedra, can seek to optimize their interactions. These additional features allow the system to exhibit approximately zero phason strain. 

One important lesson from the icosahedral glass models was that global orientational order can lead to IQC-like order (albeit perhaps not with sharp Bragg diffraction peaks). Our particle-based models go further and show that directional interactions that locally favour a given global orientational order (in this case $I_h$ symmetry), but without a second length scale, can be sufficient to generate long-range quasiperiodic order.

Although not actually proven here, given their competitive energetics and their additional entropy compared to a crystal, the assembled IQCs are likely to be thermodynamically stable over a wide temperature range. Random-tiling models \cite{Henley91} are the archetypal models for entropically stabilized quasicrystals, where the entropy of the random tiling comes from the many different ways that the tiles can be combined with negligible energy cost. However, our particle-based IQCs cannot be simply mapped onto a small set of particle-decorated tiles due to the other types of disorder that are present, and these additional sources of entropy are likely to further stabilize the IQCs.

A pertinent question is whether these models can be translated into an experimentally realizable system. Despite advances in the synthesis of patchy colloids \cite{Li20}, producing particles with the required patch geometry and selectivity is probably beyond current capabilities. Both features can be more easily controlled using DNA nanotechnology, a field that has experienced huge advances in recent years \cite{Linko22}. Specific interactions between DNA origami can be induced on the basis of shape complementarity \cite{Gerling15} or through the association of single-stranded ``sticky ends''. This strategy has already been exploited to assemble finite clusters \cite{Sigl21}, one-dimensional assemblies \cite{Hayakawa22} and even three-dimensional crystals \cite{Tian20,Liu24,Posnjak24}. 
We envision that one possible route to build our particle designs would be to use DNA origami icosahedra \cite{Liu24,Veneziano16} with single-strands to mediate edge-to-edge bonding between the relevant seven edges, thus mimicking the 7P particles. This would be quite similar to Ref.\ \citenum{Liu24} where pyrochlore crystals were grown using DNA origami icosahedra but where the bonding was between selected vertices.
Another possible route to obtain experimental analogues of our patchy particle designs is to leverage the recent advances in computational protein design \cite{Winnifrith24}.
For example, a general approach to program proteins to assemble into crystals with predefined symmetries has been recently developed \cite{Li23}. The higher symmetry of the 10P particles make them a more attractive target for protein design. Further details concerning potential experimental realization are given in Section S3 in the Supplementary Information. 

Being able to experimentally assemble an icosahedral quasicrystal from nanoparticles is an appealing goal in itself and also for practical applications. Given the high point group symmetry of IQCs, if assembled at the appropriate length scales, these structures can exhibit a small spherically symmetric optical band gap\cite{Man05}. Notably, it has been recently shown that the suitable length scales can be achieved for the assembly of a diamond lattice from DNA origami particles\cite{Posnjak24}. In light of these recent experimental advances, the prospects of
experimentally achieving a photonic icosahedral quasicrystal through directional bonds seem promising.

\section{Methods}

\subsection{Ideal quasicrystals} As target structure, we used an ideal FCI QC
generated with the cut-and-project method \cite{Steurer09}. Aperiodic structures can be produced by projecting a cut of a higher dimensional
periodic lattice onto an hyperplane that crosses the lattice with an
irrational slope. The higher dimensional space is divided into two
orthogonal subspaces (the parallel and perpendicular spaces). Aperiodic structures can be produced by projecting
the nodes of the higher dimensional lattice that fall within a region
of perpendicular space (or occupation domain) onto the parallel
space. To obtain an icosahedral quasicrystal we need to use a
six-dimensional lattice, which is the lowest dimension in which it is
possible to have a periodic lattice with icosahedral symmetry. Indeed, there are three
lattices that fulfil this requirement: the primitive, the body-centred
and the face-centred hypercubic lattices. In our previous work \cite{Noya21},
we used as target structures IQC obtained from projection of the
primitive and body-centred hypercubic lattices. Here, we generate
an ideal IQC by projection of a face-centred hypercubic lattice using
the canonical occupation domain (i.e., using the projection of the
6D unit cell onto perpendicular space as the occupation domain).
Using the same convention as in Ref.\ \citenum{Engel15}, the basis matrix is given
by:
\begin{eqnarray}
\label{eq:basis}
    Q & = & \frac{1}{a\sqrt{2 \tau^2 +2 }}
    \begin{bmatrix}
    0 & \tau & 1 & 0 & -\tau & 1 \\
    \tau & 1 & 0 & -\tau & 1  & 0 \\
    1 & 0 & \tau & 1 & 0 & -\tau \\
    0 & -1 & \tau & 0 & 1 & \tau \\
    -1 & \tau & 0 & 1 & \tau & 0 \\
    \tau & 0 & -1 & \tau & 0 & 1 \\
\end{bmatrix} \\ 
& = & (\vec{d}_1\,\vec{d}_2\,\vec{d}_3\,\vec{d}_4\,\vec{d}_5\,\vec{d}_6)
\end{eqnarray}
\sloppy The canonical occupation domain is a triacontrahedron with a circumsphere of radius $\sqrt{3\tau+2}/
(a\,\sqrt{2\tau^2+2})$. Numerical errors that
arise when the projections of some points lie exactly at the boundary
conditions can be avoided using an algorithm based on integer numbers,
as described in Ref.\ \citenum{Vogg96}. Note that this problem can also be avoided
by applying a shift to perpendicular space and leads to equivalent
structures (i.e.\ belonging to the same local isomorphism class
\cite{Steurer09}). The canonical FCI QC tiling obtained using this procedure
is depicted in Figure \ref{fig:FCI_ideal}.

Besides the FCI QC we also generated some of the lower order
rational approximants using the shear method \cite{Steurer09}. These periodic structures can be obtained by changing the irrational cut of the hyperspace by a rational one. This change of slope can also be seen as a shear transformation of the lattice in the higher dimensional space, defined by the $6\times 6$ $A^\perp$ matrix. The rational slopes are chosen as the quotient
between two consecutive integer numbers of the Fibonacci series
$(p, q)$, which are then used to designate the order of the approximant.
Thus, a lattice position of the 6D lattice, $\mathbf{r}$, is transformed according
to:
\begin{equation}
\mathbf{r'} = A^\perp \mathbf{r} 
\end{equation}
Choosing the approximant basis vectors to coincide with the cubic axes pointing to $p\,(\vec{d}_3+\vec{d}_6)\,+\,q\,(\vec{d}_2-\vec{d}_5)$, $p\,(\vec{d}_2+\vec{d}_5)\,+\,q\,(\vec{d}_1-\vec{d}_4)$ and $p\,(\vec{d}_1+\vec{d}_4)\,+\,q\,(\vec{d}_3-\vec{d}_6)$, and imposing the condition that their components in perpendicular space vanish, the shear matrix is given by:
\begin{equation}
A^\perp = 
\begin{bmatrix}
1 & 0 & 0 & 0 & 0 & 0 \\
0 & 1 & 0 & 0 & 0 & 0 \\
0 & 0 & 1 & 0 & 0 & 0 \\
A & 0 & 0 & 1 & 0 & 0 \\
0 & A & 0 & 0 & 1 & 0 \\
0 & 0 & A & 0 & 0 & 1 \\
\end{bmatrix}
\end{equation}
with $A=(q-\tau\,p)/(p+\tau\,q)$.
The lattice parameters of the $p$,$q$-approximants are given by:
\begin{equation}
a_{p,q} = \frac{2 \,(p + \tau q)}{a\,\sqrt{2\,\tau^2+2}} 
\end{equation}
It can be easily shown that this transformation is equivalent to replacing 1 by $p$ and $\tau$ by $q$ in the last three rows of the basis matrix (Eq.\ \ref{eq:basis}). Using this transformation we obtain a series of hexagonal approximants belonging to space group R3 (number 146 in the crystallographic tables).  Among these, we have chosen to perform simulations for the 3/2 approximant. The structure considered was obtained by applying a shift of (0.4, 0.4, 0.4) in perpendicular space, as this leads to a slightly higher average coordination number.  The unit cell of the 3/2 approximant generated contains 288 particles, 285 if zero-coordinated particles are disregarded. 

\subsection{Simulations}

The assembly behaviour of the model systems was investigated by Monte Carlo simulations in the canonical ensemble. Simulations were performed with a bespoke GPU parallel Monte Carlo code, using a checker board algorithm \cite{Anderson13}. The assembly behaviour was initially explored using a cubic simulation box with 20\,000 particles at a density $\rho=0.10\,\sigma_{LJ}^3$. Simulations were carried out at a temperature at which nucleation occurred via a single solid seed in reasonable simulation times (typically less than 1-3 million MC cycles, where one MC cycle amounts to an average of one move attempt per particle). In these conditions, a roughly spherical solid cluster forms in coexistence with gas. In this way, the formation of defects associated with the incommensurability of crystal formed with the cubic periodic box is avoided. Once a sufficiently big solid cluster was obtained (typically containing 15\,000-18\,000 particles), it was inserted in a larger simulation box containing a low density fluid so that the total number of particles is about 100\,000 particles. Simulations in these larger systems are performed at a higher temperature to avoid the nucleation of additional solid clusters in the fluid. The system was then allowed to equilibrate until solid-vapour equilibrium is reached, after which another 6-8 million MC cycles were performed to take averages.

Besides the assembly simulations, we also studied the thermodynamic behaviour of particle-based models of the ideal FCI QC and 3/2 approximant. The initial particle orientations in these configurations were obtained using
a Monte Carlo code that for each particle searches for the particle's orientation that provides the best alignment of its patch vectors with the interparticle vectors to its first coordination shell. Spherical clusters with about 80\,000 particles and a radius of about 30\,$\sigma$ were inserted in a cubic simulation box of edge 100\,$\sigma$ surrounded by a gas of particles with a number density of about $\rho^*=$0.02 to mimic the equilibrium conditions found for the assembled FCI QCs. These configurations were progressively heated from $T^*=0.01$ up to $T^*=0.154$ during the course of a one million Monte Carlo cycles simulation, and then equilibrated for at least 4 million MC cycles. These systems were modelled with the {7P} design and all patch-patch interactions having the same strength, $\epsilon_{AA}=\epsilon_{BB}=\epsilon_{AB}=1$.

\subsection{Analysis} 
 
 Particles belonging to the solid clusters were identified using a cluster search algorithm \cite{Rapaport04_book} with the convention that two particles are nearest neighbours if the energy between them is lower than -0.2$\epsilon_\mathrm{LJ}$. The structure of these solid clusters was characterized by evaluating the pair distribution function, the radial density and energy profiles, the coordination number distribution and the bond orientational order diagram (BOOD). The BOOD is a plot of the first coordination shell of each particle on a unit sphere, which is subsequently projected in the plane using an area-preserving Lambert projection. The single-particle dynamics was studied by evaluating the Van Hove autocorrelation function, measured after 1 million MC cycles:
\begin{equation}
G(\vec{r},t) = 1/N \sum_{i=1}^N \langle (\vec{r}_i-\vec{r}_{i,0} \rangle
\end{equation}
where $\vec{r}_{i,0}$ and $\vec{r}_{i}$ are, respectively, the initial and current positions of particle $i$.
To avoid surface effects, only the inner part of the cluster (i.e.\ particles within a distance of 25$\,\sigma_\mathrm{LJ}$ of the cluster center of mass) is considered in the calculations. 

The long range orientational order was confirmed by calculating the diffraction patterns and plotting projections along the 2-fold, 3-fold and 5-fold symmetry axes. The diffraction pattern was calculated as:
\begin{equation}
S(\vec{q})=\frac{1}{N} \left\langle \left|\sum_{i=1}^N \exp{(-i\vec{q}\vec{r}_i)}\right|^2 \right\rangle
\end{equation}
where $\vec{q}$ is the wave-vector. The diffraction pattern was averaged over 200 configurations.

The phason strain accounts for distortions of the quasiperiodic lattice in perpendicular space. Here, it was estimated from the slope of the distance between two particles in perpendicular space as a function of distance in parallel space \cite{Steurer09,Engel15}. The positions of the particles were mapped onto lattice points in the 6D space using the lifting procedure described in the Supplementary Material.

The simulation and all the analysis codes are available at (\texttt{https://github.com/evanoya/MC\_GPU}).

\section{Associated Content}

This manuscript has also been submitted to the arXiv server: Noya, E.G.; Doye, J.P.K., A one-component patchy-particle icosahedral quasicrystal. 2024, arXiv. https://arxiv.org/abs/2407.17212 (July 24th, 2024).

\begin{acknowledgement}
EGN acknowledges finantial support by grants PID2020-115722GB-C21 and PID2023-151751NB-I00 funded by MCIN/AEI/10.13039/501100011033. 
Computer resources and technical assistance provided by CESGA (Centro de Supercomputación de Galicia) are also gratefully acknowledged.

\end{acknowledgement}


\begin{suppinfo}

Further model design, methods and analysis of the assembled quasi-crystals is provided as Supporting Information.
Supplementary Movies M1 and M2 show the nucleation and growth of the FCI QCs at ($\epsilon_{BB}/\epsilon_{AA}=1.2$) and above ($\epsilon_{BB}/\epsilon_{AA}=1.4$) the transition from a gas of particles to a gas of icosahedra in the fluid phase.

\end{suppinfo}

\providecommand{\latin}[1]{#1}
\makeatletter
\providecommand{\doi}
  {\begingroup\let\do\@makeother\dospecials
  \catcode`\{=1 \catcode`\}=2 \doi@aux}
\providecommand{\doi@aux}[1]{\endgroup\texttt{#1}}
\makeatother
\providecommand*\mcitethebibliography{\thebibliography}
\csname @ifundefined\endcsname{endmcitethebibliography}
  {\let\endmcitethebibliography\endthebibliography}{}

\end{document}


\renewcommand{\thefigure}{S\arabic{figure}}
\renewcommand{\theequation}{S\arabic{equation}}
\renewcommand{\thetable}{S\arabic{table}}
\renewcommand{\thesection}{S\arabic{section}}

\section{Model design and methods}

\subsection{Environment analysis}

The first step to design a model system is to analyse the local environments in the target FCI QC. As can be seen in the radial distribution function of the ideal FCI QC (Fig.\ \ref{fig:structure}A), the first and second neighbours give rise to two well-separated peaks. In the local environment analysis, only the first neighbours are considered. The thirteen local environments, with between zero and seven neighbours, found in the ideal FCI QC are reported in Table\ \ref{tbl:local_environments_FCI}. The angles between bonds in each local environment, as well as the proportion in which each local environment appears are also provided. The local environments with two or more neighbours are also depicted in Fig\ 1D in the main text. 
As can be seen in Table\ \ref{tbl:local_environments_FCI}, the most frequent environments in the ideal FCI QC are also quite common both in the trigonal 3/2 and 5/3 approximants. However, some of the less common environments in the ideal FCI QC are not present in these approximants. Similarly, the six-coordinate environment encountered (although in a very low proportion) in the two approximants and labeled as 12b in Table\ \ref{tbl:local_environments_FCI}, is not present in the ideal FCI QC. 

\begin{table}[t]\centering
\caption{\label{tbl:local_environments_FCI} \textnormal{Local environments in the ideal FCI QC. The number of nearest neighbours is defined as the number of particles within a distance of 1.3\,$\sigma$ of the bond distance (this cutoff provides a clear distinction between first and second coordination shells, see Fig.\ \ref{fig:structure}). The angles formed between a particle and its pairs of nearest neighbours are provided in the third column. The proportion with which each local environment appears in the ideal FCI QC is given in the fourth column. The environment labelled as 12b appears in the approximants but not in the QC.}}
\scriptsize
\begin{tabular}{lcccccccccc}

\hline
Index  & & $n_\mathrm{neigh}$ && Angles   && FCI QC && 3/2 approximant && 5/3 approximant \\
\hline
1  & &  0   & &   & &  0.013 & & 3/288=0.010 && 16/1220=0.013  \\
2  & &  1   & &   & & 0.003 && 3/288=0.010 && 3/1220=0.002 \\
3  & &  2   & &  108.00  & &  0.004 & &  -- && -- \\
4  & &  3   & &  108.00$\times$2, 60.00   & &   0.002  && -- &&  3/1220=0.002 \\
5  & &  3  & &  108.00 $\times$3  & &  0.007 && 1/288=0.003 &&  7/1220=0.006 \\
6  & &  4   & & 108.00$\times$3, 60.00$\times$3 & &  0.001 && -- &&  --  \\
7  & &  4   & & 144.00, 108.00$\times$4, 60.00   & &  0.004  && 3/288=0.010 && 6/1220=0.005 \\
8  & &  5   & &  144.00$\times$2, 108.00$\times$6, 60.00$\times$2  & &  0.066  && 21/288=0.073 && 84/1220=0.069 \\
9  & &  5   & &  108.00$\times$5, 60.00$\times$5  & &  0.003  && -- && 3/1220=0.002 \\
10 & &  5   & & 144.00$\times$2, 108.00$\times$5, 60.00$\times$3   & &  0.003  && -- &&  24/1220=0.020 \\
11 & &  6   & & 144.00$\times$2, 108.00$\times$7, 90.00, 60.00$\times$5   & &  0.003 && 3/288=0.010 &&  6/1220=0.005 \\
12 & &  6   & & 144.00$\times$3, 108.00$\times$7, 90.00, 60.00$\times$4 & & 0.205  && 57/288=0.198 && 243/1220=0.199 \\
12b & &  6   & & 144.00$\times$3, 108.00$\times$6, 90.00$\times$3, 60.00$\times$3 & & -- & & 2/288=0.007 && 3/1220=0.002 \\
13 & &  7   & & 144.00$\times$4, 108.00$\times$9, 90.00$\times$2,  60.00$\times$6  & &  0.686 && 195/288=0.678 && 837/1220=0.686 \\
\hline
\end{tabular}
\end{table}

\subsection{Particle designs}

The thirteen local environments found in the ideal FCI QC display a common geometry. All can be represented by a patchy particle mimicking the highest coordinate environment, the one with seven neighbours, if we allow for the possibility that not all the patches always form bonds. Our first model, named {7P} FCI, consists of only one particle type with seven patches pointing towards the first neighbours in this environment. The patch vectors, as well as their reference vectors and offset angles for the torsion term, are provided in Table\ \ref{tbl:7p_fci_model}. Patches are classified into two types: five patches of type A, which often form intra-icosahedral bonds, and  two patches of type B, which are mostly involved in inter-icosahedral bonds. We explored the relative interaction strength of the patch-patch interactions $\epsilon_{BB}/\epsilon_{AA}$ (taking $\epsilon_{AA}$=1 and $\epsilon_{AB}=(\epsilon_{AA}+\epsilon_{BB})/2$), finding that in our simulations assembly is only successful when $\epsilon_{BB}/\epsilon_{AA} > 1.2$.

Aiming at finding a simpler model system in which all the patch-patch interactions are equal, we propose a second model in which particles carry five equivalent $B$ patches (instead of only two), named as {10P} FCI model. Due to geometric constraints the maximum number of bonds is still limited to seven (although eight weaker bonds can be also found in transient configurations). Owing to the increased configurational entropy associated with the bonding of type B patches, this model is able to assemble into an FCI QC when all patch-patch interactions are of equal strenght. The geometric properties of the patches for this model are given in Table\ \ref{tbl:10p_fci_model}.

The {10P} model can be further simplified by replacing the five $A$ patches by a single ring patch located at a cone angle $\mu_A =\arccos{(\tau/{\sqrt{4\,\tau+3}})}=$1.017 rad= 58.28$^{\circ}$ with respect to one pole of the particle, and the five $B$ patches by a second ring patch at a cone angle $\mu_B = \arccos{(\tau/\sqrt{2+\tau})}$=0.5536 rad= 31.17$^{\circ}$ with respect to the opposite pole. The rings are thus represented by two patches pointing at opposite poles of the spherical particles and the angular term in the potential is modified to \cite{Wilber07}:
\begin{equation}
    V_\mathrm{ang,ring}(\mathbf{\hat{r}}_{ij},\mathbf{\Omega}_i,\mathbf{\Omega}_j) = \exp \left( -\frac{(\theta_{\alpha ij}-\mu_{\alpha})^{2}}{2\sigma_{\mathrm{ang,ring}}^{2}} \right) \exp \left( -\frac{(\theta_{\beta ji}-\mu_{\beta})^{2}}{2\sigma_{\mathrm{ang,ring}}^{2}} \right) .
    \label{eq:Vang_ring}
\end{equation}
$\theta_{\alpha ij}$ is the angle between the patch vector $\mathbf{\hat{P}}_{i}^{\alpha}$, representing the patch $\alpha$, and $\mathbf{\hat{r}}_{ij}$. $\sigma_\mathrm{ang,ring}$ is a measure of
the angular width of the ring patch. We set $\sigma_\mathrm{ang,ring}$=0.15 rad to reduce the surface area covered by the patches and prevent the appearance of a liquid phase. The torsional dependent term was not included for particles with ring patches. This model is named {2R} FCI.

\begin{figure}
\centering
\includegraphics[width=15cm]{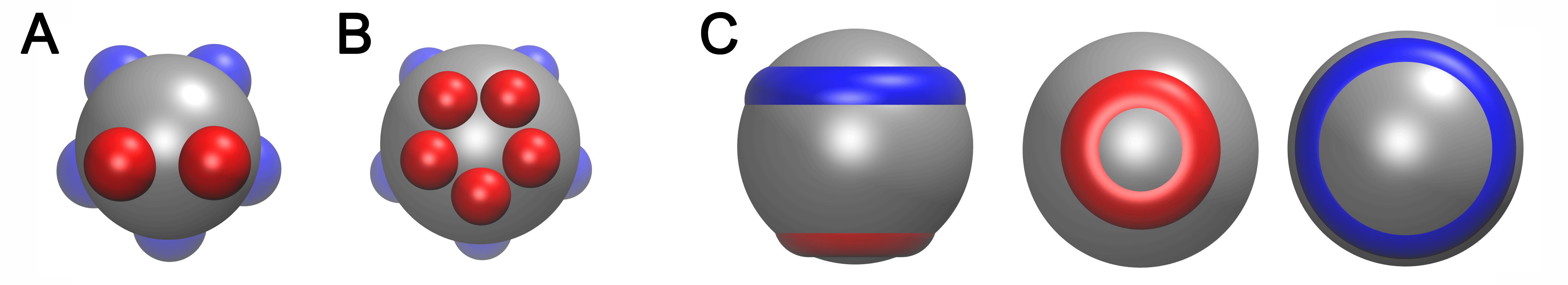}
\caption{\label{Fig:particle_types} Particle designs studied in this work: ($A$) {7P} , ($B$) {10P} model and ($C$) {2R} models (in $C$, from left to right: front, top and bottom views are depicted to ease visualization).}
\end{figure}

These three particle designs are depicted in Fig.\ \ref{Fig:particle_types}.

\begin{table}\centering
\caption{\label{tbl:7p_fci_model} \textnormal{Description of the geometry of the patchy particle used to assemble the {7P FCI} QC. The positions of the patches on the particle surface are specified by the patch unit vectors. For each patch, the reference vector and offset angles used for evaluating the torsional interactions are provided. Patches are divided into two types: type A comprises patches 1 to 5 that often form intra-icosahedral bonds, and type B includes patches 6 and 7 that most commonly form inter-icosahedral bonds. All the patches can interact with any other patch. The strength of the intra-icosahedral bonds is set to 1 ($\epsilon_{AA}=1$), and an arithmetic mixing rule is used to calculate the cross interactions between inter- and intra-icosahedral patches ($\epsilon_{AB}=\epsilon_{AA}+\epsilon_{BB})/2$); thus, there is only adjustable parameter $\epsilon_{BB}/\epsilon_{AA}$.}}
\scriptsize
\begin{tabular}{ccccccccc}
\hline
Patch number & & Patch type &&Patch vector & & Reference vector & & Offset angles  \\
 %
 %
 %
 %
 %
 %

 \hline
 P$^1$    & &  A  && (0.5,  0.30901699, -0.80901699) & &  (0, 0.52573111, 0.85065081)  & & 0, 180  \\
 P$^2$    & &  A && (0.80901699, -0.5, -0.30901699)  & & (0, 0.52573111, 0.85065081) & & 0, 180  \\
 P$^3$    & &  A && (0, -1, 0)  & & (0, 0.52573111, 0.85065081)  & & 0, 180   \\
 P$^4$    & &  A && (-0.80901699, -0.5, -0.30901699) & & (0, 0.52573111, 0.85065081) & & 0, 180  \\
 P$^5$    & &  A &&  (-0.5, 0.30901699, -0.80901699)  & & (0, 0.52573111, 0.85065081) & & 0, 180  \\
 P$^6$    & &  B &&  (0, 0, 1)  & & (0.85065807, 0, 0.52573111) & & 0, 180   \\
 P$^7$    & &  B && (0.30901733, 0.80901699, 0.5) & &  (0.85065807, 0, 0.52573111) & & 0, 180  \\
\hline
\end{tabular}

\end{table}

\begin{table}\centering
\caption{\label{tbl:10p_fci_model} \textnormal{Description of the geometry of the patchy particle used to assemble the {10P FCI} QC. The positions of the patches on the particle surface are specified by the patch unit vectors. For each patch, the reference vector and offset angles used for evaluating the torsional interactions are provided. Patches are divided in two types: type A comprises patches 1 to 5 and often form intra-icosahedral bonds, and type B includes patches 6 to 10 that most commonly form inter-icosahedral bonds. All the patches can interact with any other patch with the same strength, which is set $\epsilon_{AA}=\epsilon_{BB}=\epsilon_{AB}=1$. The width of the patches is set to $\sigma_{ang}=0.25$\ rad when the torsional contribution to the energy is not included and to $\sigma_{ang}=0.30$\ rad for the model with torsions. Narrower patches are required in the model without torsions to avoid the presence of a liquid phase (see main text for a more detailed explanation).}}
\scriptsize
\begin{tabular}{ccccccccc}
\hline
Patch number & & Patch type &&Patch vector & & Reference vector & & Offset angles  \\
\hline
 P$^1$    & &  A  && (0.5,  0.30901699, -0.80901699) & &  (0, 0.52573111, 0.85065081)  & & 0, 180  \\
 P$^2$    & &  A && (0.80901699, -0.5, -0.30901699)  & & (0, 0.52573111, 0.85065081) & & 0, 180  \\
 P$^3$    & &  A && (0, -1, 0)  & & (0, 0.52573111, 0.85065081)  & & 0, 180   \\
 P$^4$    & &  A && (-0.80901699, -0.5, -0.30901699) & & (0, 0.52573111, 0.85065081) & & 0, 180  \\
 P$^5$    & &  A &&  (-0.5, 0.30901699, -0.80901699)  & & (0, 0.52573111, 0.85065081) & & 0, 180  \\
 P$^6$    & &  B &&  (0, 0, 1)  & & (0.85065807, 0, 0.52573111) & & 0, 180   \\
 P$^7$    & &  B && (0.5, 0.30901733, 0.80901699) & &  (0.52573111, 0.85065807, 0) & & 0, 180  \\
 P$^8$    & &  B && (-0.5, 0.30901733, 0.80901699) & &  (0, -0.52573111, 0.85065807) & & 0, 180  \\
 P$^9$    & &  B && (0.30901733, 0.80901699, 0.5) & &  (-0.52573111, 0.85065807, 0) & & 0, 180  \\
 P$^{10}$    & &  B && (-0.30901733, 0.80901699, 0.5) & &  (-0.85065807, 0, 0.52573111) & & 0, 180  \\
\hline
\end{tabular}

\end{table}

\subsection{Generalizability of approach}

Here we have developed patchy-particle systems to form a FCI QC. This complements previous work that used as design targets IQCs that were generated from body-centred and primitive hypercubic lattices \cite{Noya21}. In these three cases, only the lattice points associated with the hypercubic lattices were projected onto the physical space. In principle, more target IQCs could be generated by projecting from hypercubic lattices using different occupation domains or with particles placed at additional positions in the hypercubic unit cell as well as the lattice points. How best to perform this in a way that generates target IQCs with interparticle distances that can be realized with particles with well defined radii is less clear and a trial-and-error approach might be necessary. Experimental IQCs for which the structure has been determined could also be used as a source for target IQCs for our patchy-particle design procedure. However, patchy particles are generally less well-suited for realizing structures with high coordination numbers such as are typical in metallic alloy IQCs.

IQCs are the only QCs that are both quasiperiodic in all 3-dimensions and possess non-crystallographic symmetry. Our approach relies on the incompatibility of the particle's local patch geometry and interactions with any periodic crystalline symmetry to generate quasiperiodic order and so it is unlikely to be possible to use patchy particles to generate quasicrystals with crystallographic symmetries \cite{Lifshitz02}. 

Our general approach can also be used to generate patchy-particle systems that can form 3-dimensional axial QCs, i.e.\ QCs with quasiperiodic order in two dimensions and periodic order in the third. Examples with dodecagonal \cite{vanderLinden12,Reinhardt16} and, very recently, octagonal \cite{Kowaguchi24} order have been reported. However, generating suitable target axial QCs is not straightforward, as although projection from higher-dimensional can be straightforwardly used to generate 2-dimensional quasiperiodic structures, how to best order the particles in the third dimension is less clear. Despite this, we hope that our general approach might not only be used to generate quasicrystals with all the symmetries observed experimentally, but also with symmetries not yet realized in any material, e.g.\ a heptagonal QC.

\subsection{Lifting}

``Lifting'' involves mapping the positions of particles in an IQC configuration to lattice sites in a 6D hypercubic lattice. In effect it is the inverse of the cut-and-project method that was used to generate the ideal IQCs. The basic procedure we used is as follows. First, one starts at an arbitrary particle (we usually choose the particle closest to the centre of mass of a cluster) that is assigned to the origin of the 6D lattice. One then follows the bond network away from this particle assigning 6D lattice sites to its neighbours. This procedure is then iterated from the newly-assigned particles until all particles have been lifted. 

\begin{figure}[t]
\centering
\includegraphics[width=9cm]{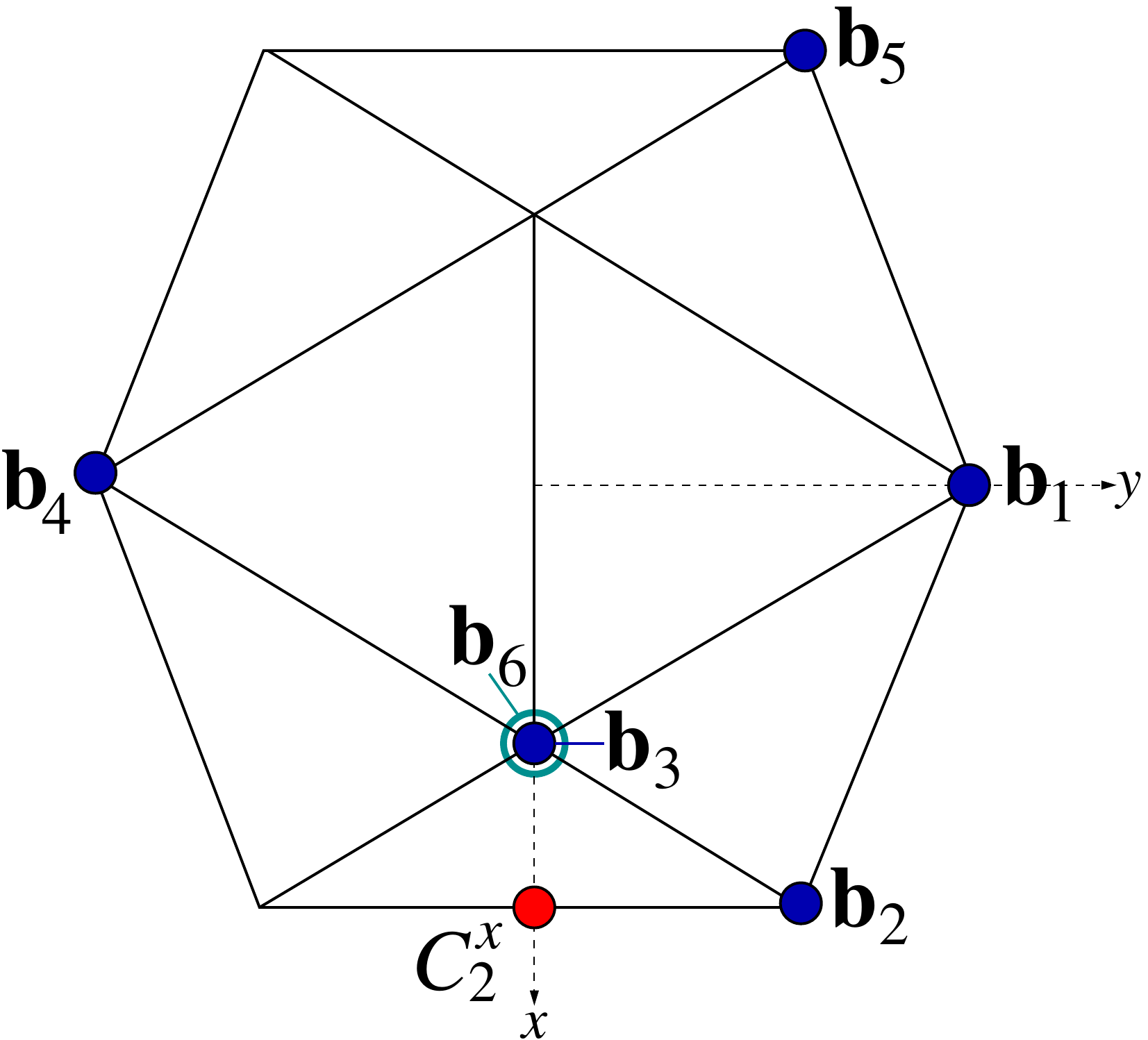}
\caption{The $\mathbf{b}_i$ vectors obtained by projection of the unit vectors along the 6 dimensions of the hypercubic lattice into parallel space and their relationship to the five-fold symmetry axes of an icosahedron that is oriented with three of its two-fold symmetry axes along $x$, $y$ and $z$. Note that the cyan ring associated with $\mathbf{b}_6$ is to signify it has a negative component in $z$.}
\label{fig:bvectors}
\end{figure}

To achieve this lifting one needs a mapping between a bond vector in a given direction and an inter-lattice vector
in 6D. In our case the bonds are just along the 30 directions of the $C_2$ symmetry axes of the $I_h$ point group.
For example, vectors along the $C_2$ axis in the $+x$ direction can be generated by $\mathbf{b}_2-\mathbf{b}_5$ (i.e.\ a sum of the five-fold vectors that pass through the vertices at either end of the edge of the icosahedron that the symmetry axis passes through) or $\mathbf{b}_3+\mathbf{b}_6$ (i.e.\ a sum of the five-fold vectors that pass through the vertices at the opposite corners of the triangular faces containing the edge that the symmetry axis passes through), or any linear combinations thereof (Fig.\ \ref{fig:bvectors}). Note $|\mathbf{b}_2-\mathbf{b}_5|=\tau|\mathbf{b}_3+\mathbf{b}_6|$. Thus, in general, for a bond along the $C_2$ direction in $x$, the corresponding step in the 6-dimensional hypercubic lattice will be of the form: 
\begin{equation}
\Delta\mathbf{l}=\frac{a}{2}\left(0,c,d,0,-c,d\right)
\label{eq:deltal_gen}
\end{equation}
where $c$ and $d$ are integers.
In our case, comparison of  bonds in the ideal FCI QC with the lattice points from which they were projected allows us to determine values of $c$ and $d$ consistent with this projection. Thus we obtain
\begin{equation}
\Delta\mathbf{l}=\frac{a}{2}\left(0,2,1,0,-2,1\right).
\end{equation}
Equivalent 6D displacement vectors for the other 29 bond directions can easily be derived. 
The length of the bond in the ideal FCI QC in terms of $a$ is
\begin{equation}
r_0=|\mathbf{Q}_\mathrm{par}\Delta\mathbf{l}^T|=\frac{1+ 2 \tau}{\sqrt{4+2\tau}} a .
\label{eq:real_to_hyper}
\end{equation}

If, as is the case for the current system, only bonds along the 2-fold axes are present, then the lifting is guaranteed to result in a face-centred hypercubic lattice. If the lattice sites are denoted by 
\begin{equation}
\frac{a}{2}\left(n_1,n_2,n_3,n_4,n_5,n_6\right),
\end{equation}
where the $n_i$ are integers, the definition of a face-centred lattice is that the $\sum n_i$ is even. It is easy to see that the change in this sum associated with the interlattice vector defined in Eq.\ \ref{eq:deltal_gen} must also be even. Hence, lifting can only then generate lattice sites consistent with a face-centred lattice.

In practice, bonds in our simulated configurations will not exactly point along a given symmetry direction. We instead require that the dot product of the normalized bond vector with the unit vector along the symmetry direction must be greater than 0.98 to trigger lifting, adopting the convention that two particles are bonded if the distance between them is lower than 1.35\,$\sigma$. Results are fairly similar for threshold values of the dot product within 0.97-0.99 and for bond cutoffs within 1.3-1.4\,$\sigma$. We also exclude any particles from the analysis where ambiguous assignments arise due to multiple paths to the same particle leading to different assignments. In our case this is mainly due to some kind of local defect and the number of such instances is reduced with a stricter cutoff for lifting. If dislocations were present, the resulting ambiguity problem would effectively render the whole lifting process ill-defined. Dislocations were not observed in any of our systems. 

As expected, the lifting process only leads to lattice sites consistent with a face-centred lattice. We also checked the occupancy of the 32 lattice sites in the primitive face-centred unit cell. They had the same occupancy to within 0.1-0.2\% further confirming the face-centred character of the assembled IQCs.

 For the primitive (PI QC) and body-centred (BCI QC) icosahedral quasicrystals, the lifting procedure is analogous to the one just described. The main difference is that in the PI QC, bonds exist along the 2-fold and 3-fold axes, whereas in the BCI QC, bonds are present along the 2-fold, 3-fold, and 5-fold axes, with different bond lengths.  In contrast, in the FCI QC all bonds are aligned with 2-fold axes. Given that the coordinates in 3D and 6D are known for the ideal QCs, mapping a bond in a given direction to an inter-lattice vector in 6D is straightforward.

\subsection{Indexing of diffraction patterns}

\begin{figure}
\centering
\includegraphics[width=16.7cm]{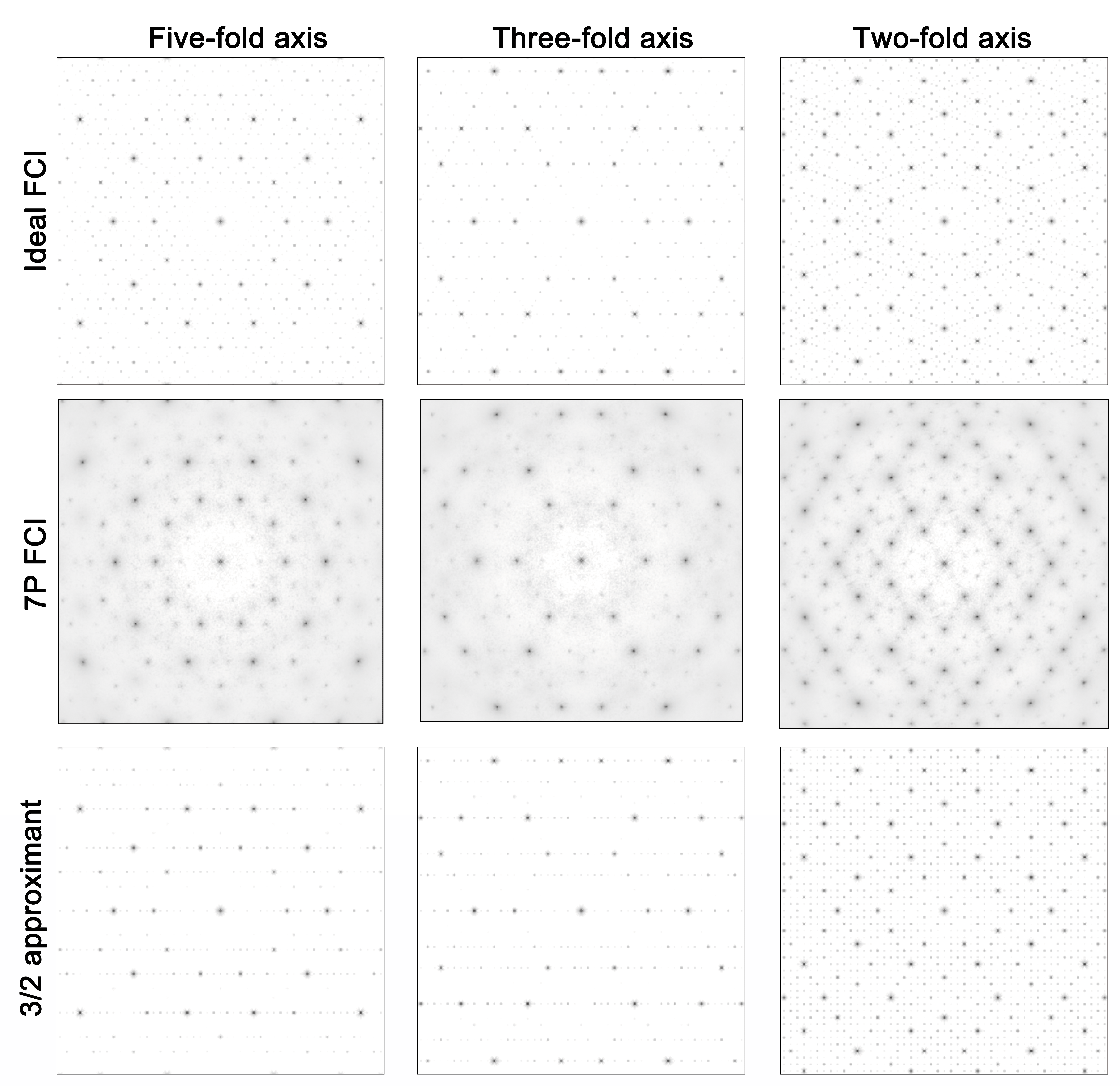}
\caption{\label{Fig:patterns_ideal_vs_simul} Diffraction pattern along the 5-fold, 3-fold and 2-fold symmetry axes of the ideal FCI QC, the assembled QC with the {7P} model and $\epsilon_{BB}/\epsilon_{AA}=1.2$. In the bottom row, projections of the diffraction pattern of the trigonal 3/2 approximant projected along the equivalent axes are also shown.}
\end{figure}

The diffraction pattern along the five-fold, three-fold and two-fold symmetry axes of the ideal FCI QC and assembled QC are shown in  
Fig.\ \ref{Fig:patterns_ideal_vs_simul}. For comparison, projections of the pattern of the trigonal 3/2 approximant along equivalent axes are also shown. It can be observed that the brightest peaks in the pattern of the ideal structure are present in the diffraction pattern of the assembled QC, again supporting that it is of FCI type. This is more rigorously confirmed by indexing of the diffraction pattern.

Peaks in the IQC diffraction patterns are expected to occur at scattering vectors corresponding to the projection of the reciprocal lattice of the 6D hypercubic lattice onto parallel space. Six indices are required to index each peak. One common indexing scheme uses the six vectors obtained by projecting the six directions of the reciprocal lattice as a basis set; these vectors are along the fivefold axes of symmetry of $I_h$. We instead choose to use the indexing scheme where two basis vectors with a length ratio of $\tau$ along each of $x$, $y$ and $z$ are used; i.e.\
\begin{equation}
    \mathbf{q}=\frac{2\pi}{a\sqrt{4+2\tau}}(h+h^\prime \tau,k+k^\prime\tau,l+l^\prime\tau).
\end{equation}
The peaks are then labelled using the notation $(h/h^\prime  k/k^\prime l/l^\prime)$.

The 6D hypercubic lattice type can be identified from the diffraction pattern because each has a different set of extinction conditions. For an FCI system all indices must be even, and $h+l+h^\prime+k^\prime$ and $h+k+l^\prime+k^\prime$ (and $l+k+h^\prime+l^\prime$) must be a multiple of 4. The differences are most apparent for the diffraction pattern along the two-fold axis.

One of the difficulties of indexing IQC diffraction patterns is that the pattern of peaks is unchanged by a scaling of $\tau$ or $\tau^3$ (depending on the symmetry direction). There is not simply a first peak that allows one to start the indexing and to work out the relationship between the real space distances and the hypercubic lattice parameter. Instead, when indexing experimental patterns the intensity of the peaks are often used to work out this relationship, as their intensity is related to the magnitude of $\mathbf{q}_\mathrm{perp}$, where $\mathbf{q}_\mathrm{perp}$ is the result of 
projecting the relevant reciprocal lattice vector onto the perpendicular space.
\begin{equation}
    \mathbf{q}_\mathrm{perp}=\frac{2\pi}{a\sqrt{4+2\tau}}[(h^\prime - h \tau)\mathbf{i}+(k^\prime-k \tau)\mathbf{j}+(l^\prime -l \tau)\mathbf{k}] 
\end{equation}
In particular, the intensity decreases with increasing $q_\mathrm{perp}$.

In our case, however, we already know the relationship between the hypercubic lattice parameter and the bond distance in the ideal quasicrystal (Eq.\ \ref{eq:real_to_hyper}), and so we can use this to index the peaks. For example, a peak at $(h/h^\prime,0/0,0/0)$ would be expected to occur on the $x$-axis at 
\begin{equation}
|\mathbf{q}|=
\frac{2\pi(h+h^\prime \tau)(1+2\tau)}{(4+2\tau) r_0}
\end{equation}
where the average bond distance $r_0$ is expected to be approximately equal to $2^{1/6}\sigma_\mathrm{LJ}$, the position of the minimum in the Lennard-Jones potential.

\begin{figure}[t]
\centering
\includegraphics[width=10cm]{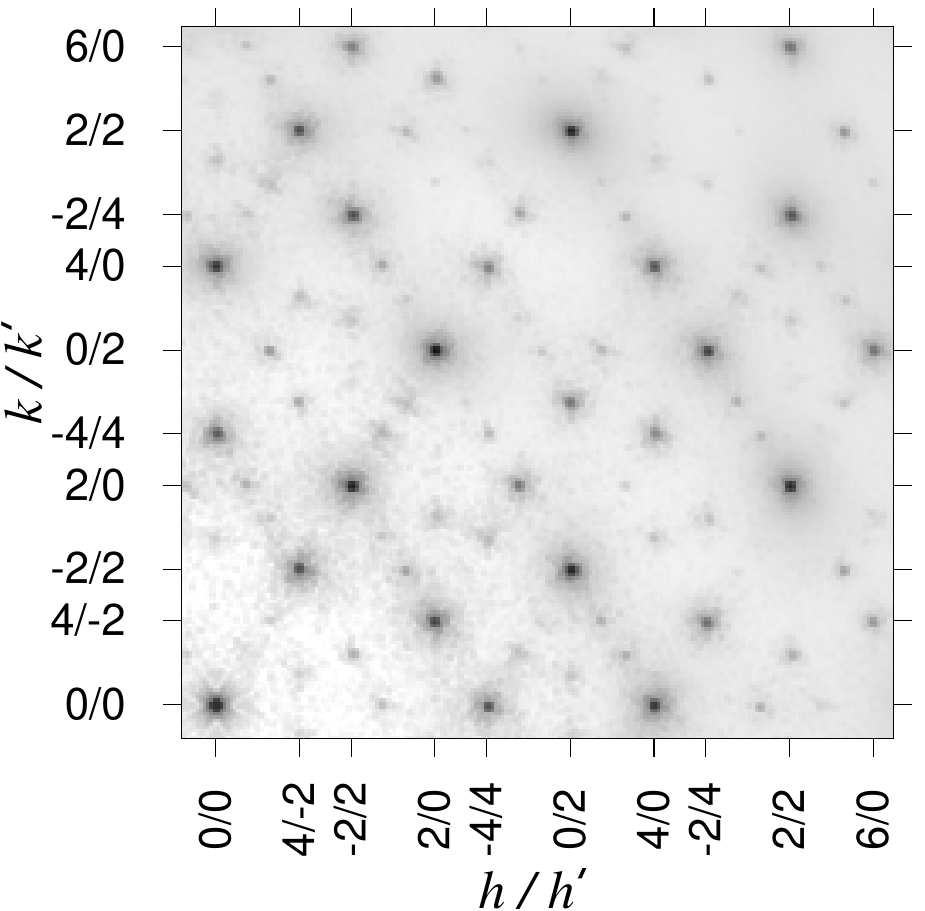}
\caption{\label{fig:indexed} Indexed diffraction pattern for the assembled IQC viewed along a two-fold axis.
}
\end{figure}

A diffraction pattern along the 2-fold axis indexed in this manner is illustrated in Fig.\ \ref{fig:indexed}. The pattern shows the distinctive features of an FCI QC with all observed peaks consistent with the extinction conditions for this lattice.

\section{Results}

\subsection{Effect of the interaction strength}

\begin{figure}[!hp]
\centering
\includegraphics[width=16.7cm]{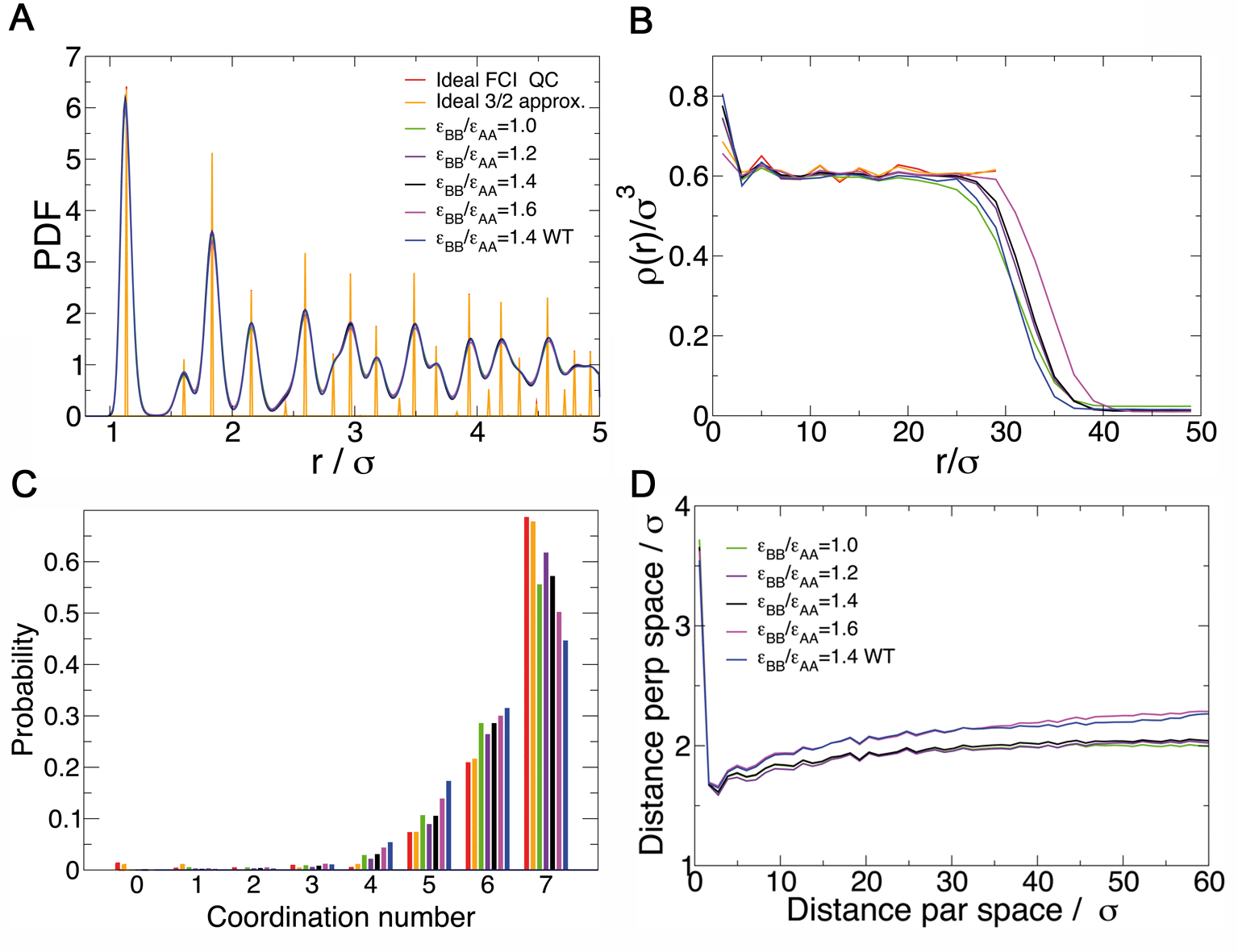}
\caption{\label{fig:structure} ($A$) Pair distribution function, ($B$) radial density, ($C$) coordination number and ($D$) phason strain of the assembled QC with the {7P} FCI model at different $\epsilon_{BB}/\epsilon_{AA}$ interaction strengths and with and without torsions, as indicated in the legends. The temperatures at which each model was simulated are given in Table ~\ref{tbl:bond_type}. Assembly of the IQC from the gas phase was not observed in our simulations of the {7P} FCI model with $\epsilon_{BB}/\epsilon_{AA}=1$. For this case, these structural properties were evaluated in a cluster grown with  $\epsilon_{BB}/\epsilon_{AA}=1.2$, subsequently simulated with $\epsilon_{BB}/\epsilon_{AA}=1.0$.  For comparison, results for the ideal FCI QC and its 3/2 approximant (without zero-coordinated particles) are also provided. The peaks of the PDF of the ideal structures have been scaled to ease comparison with the simulated PDFs. The ideal FCI QC and its trigonal 3/2 approximant exhibit very similar PDFs within the shown distance range, with peaks exactly at the same distances and only small differences in the peaks intensity.  }
\end{figure}

\begin{figure}[!hp]
\centering
\includegraphics[width=13cm]{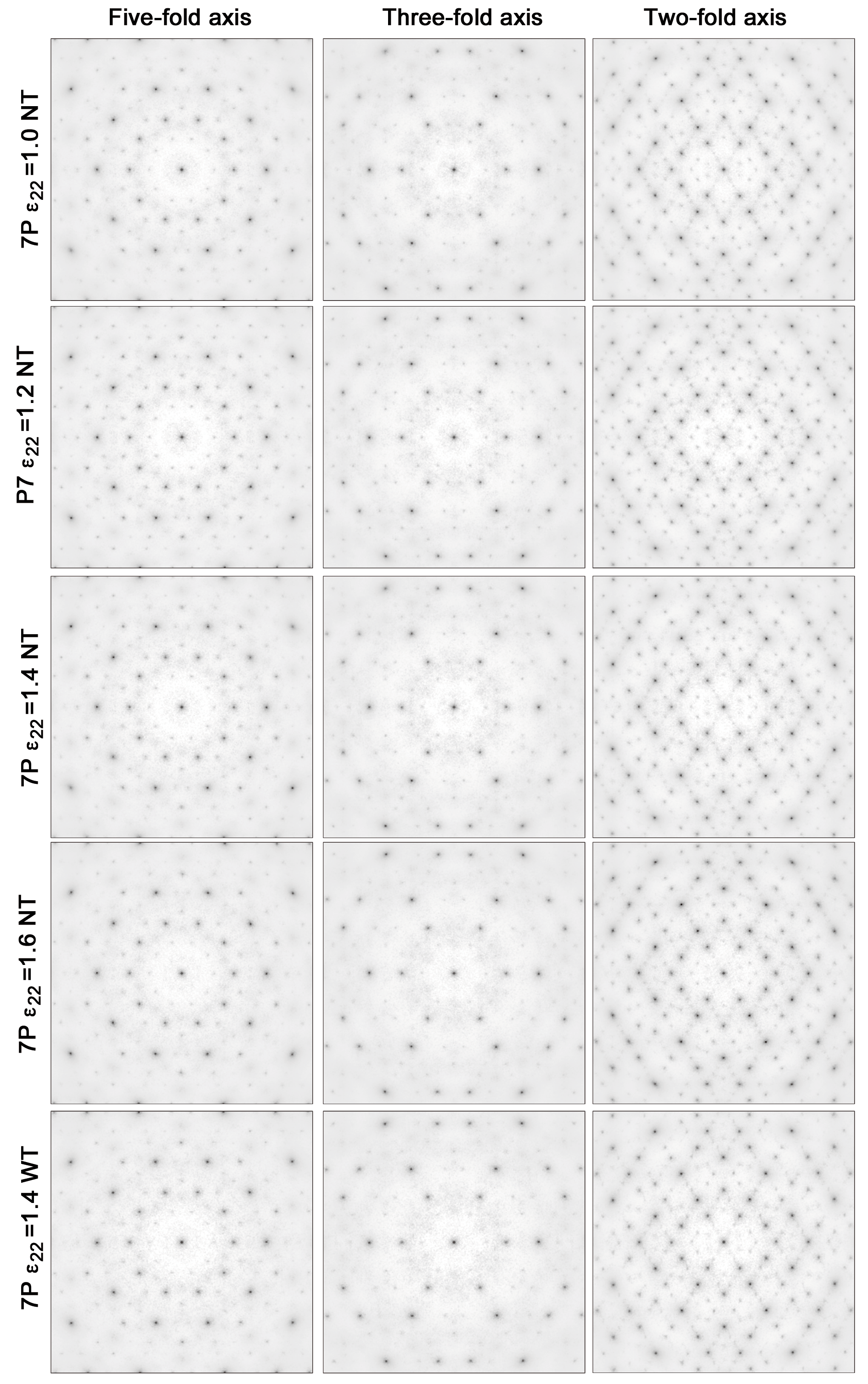}
\caption{\label{fig:pattern_vs_eps} Diffraction pattern of the assembled IQC with the {7P} FCI model for different $\epsilon_{BB}/\epsilon_{AA}$ interaction strengths and with and without torsions. The temperatures at which each model was simulated are given in Table ~\ref{tbl:bond_type}. }
\end{figure}

As discussed in the main text, simulations of the {7P} FCI model lead to the assembly of the IQC for interaction strengths within $1.2 \lesssim \epsilon_{BB}/\epsilon_{AA} \lesssim 1.8$. As can be seen in Fig.\ \ref{fig:structure}, the radial distribution function and radial density are very similar for all the interaction strengths and also  when a torsional term is added. Some differences are found in the distribution of the coordination numbers (Fig.\ \ref{fig:structure}C). All the assembled clusters exhibit a lower proportion of 7-coordinate particles than the ideal FCI QC. In the ideal FCI QC about 65\% of the particles are fully bonded, whereas in the assembled QC this percentage falls down to 45-60\%. The highest percentage is attained for $\epsilon_{BB}/\epsilon_{AA}=1.2$, and it decreases either when $\epsilon_{BB}/\epsilon_{AA}$ increases and when torsions are included. The average coordination number (Table\ \ref{tbl:bond_type}) is also similar in all the systems, but again the system with $\epsilon_{BB}/\epsilon_{AA}=1.2$ is the one with an average coordination closer to that of the ideal FCI QC. Projections of the diffraction pattern along the symmetry axes (Fig.\ \ref{fig:pattern_vs_eps}) are very similar independent of the model parameters or the presence of the torsional term. Some differences appear in the phason strain plots (Fig.\ \ref{fig:structure}D), with higher strains appearing for higher values of $\epsilon_{BB}/\epsilon_{AA}$ and when the potential contains the torsional term (Table\ \ref{tbl:phason_strain}).

\begin{table}[!hp]\centering
\caption{\textnormal{Fraction of $AA$, 
$BB$ and $AB$ bonds ($x_{AA}$, $x_{BB}$ and $x_{AB}$) and average coordination number ($\langle CN\rangle$) in the simulated IQCs with the {7P} and {10P} FCI models, along with the simulated temperature and the equilibrium cluster size ($N_{c}$). Results for the ideal FCI QC and the trigonal 3/2 approximant simulated with the {7P} FCI model and $\epsilon_{BB}/\epsilon_{AA}=1.0$ are also provided. Two particles are considered bonded if the interaction between them is lower than -0.20\,$\epsilon$. For a few cases, the average coordination number was also calculated using a distance criteria with a cutoff distance of 1.3\,$\sigma$ to define first neighbours (data within parentheses). Typically only those particles within a sphere of radius 20\,$\sigma$ from the center of mass, which are sufficiently away from the cluster surface (see Fig.\ \ref{fig:structure}B), are considered in this analysis. For comparison, the average coordination number in the ideal FCI QC and the lower approximants are given in the lower rows.}}
\begin{tabular}{lcccccc}
\toprule
System &  $T^*$ & $N_{c}$ & $x_{AA}$ & 
$x_{BB}$ & $x_{AB}$ & $\langle CN\rangle$ 
\\
\midrule
{7P}, $\sigma=0.30$, $\epsilon_{BB}=$1.0, NT & 0.154 & 80000 & 0.65 & 0.21 & 0.14  & 6.33 \\
{7P}, $\sigma=0.30$, $\epsilon_{BB}=$1.2, NT & 0.159 & 69000 & 0.65 & 0.22 & 0.13 & 6.43 (6.51) \\
{7P}, $\sigma=0.30$, $\epsilon_{BB}=$1.4, NT & 0.164 & 85000 & 0.63 & 0.22 & 0.15 & 6.37 \\
{7P}, $\sigma=0.30$, $\epsilon_{BB}=$1.6, NT & 0.170 & 100000 & 0.60 & 0.22 & 0.19  & 6.20 \\
\midrule
{7P}, $\sigma=0.30$, $\epsilon_{BB}=$1.4, WT  & 0.159 & 75000 & 0.60 & 0.19 & 0.21  & 6.10 \\
\midrule
Ideal FCI QC, {7P}, $\sigma=0.30$, $\epsilon_{BB}=$1.0, NT & 0.154 & 82000 & 0.65 & 0.23 & 0.12 &   6.44 \\
 Ideal 3/2 AC, {7P}, $\sigma=0.30$, $\epsilon_{BB}=$1.0 NT & 0.154 & 82000 & 0.65 & 0.23 & 0.12  & 6.44 \\
\midrule
{10P} $\sigma=0.25$, $\epsilon_{BB}=1$, NT & 0.151 & 77000 &  0.62 & 0.22 & 0.16 &  6.42 (6.58) \\
{10P} $\sigma=0.30$, $\epsilon_{BB}=1$, WT & 0.154 & 90000 & 0.56 & 0.20 & 0.24 & 6.25 (6.59) \\
\midrule
{2R} $\sigma=0.15$, $\epsilon_{BB}=1$ & 0.166 & 83000 & 0.61 & 0.25 & 0.14 & 6.70 (7.37) \\
\midrule
0/1 approximant & -- & -- & -- & -- & -- & 4.0 \\
1/1 approximant & -- & -- & -- & -- & -- & 6.38 \\
2/1 approximant & -- & -- & -- & -- & -- & 6.44 \\
3/2 approximant & -- & -- & -- & -- & -- & 6.45 \\
5/3 approximant & -- & -- & -- & -- & -- & 6.46 \\
8/5 approximant & -- & -- & -- & -- & -- & 6.46 \\
FCI QC  & -- & -- & -- & -- & -- & 6.47 \\

\bottomrule
\label{tbl:bond_type}
\end{tabular}
\end{table}

\begin{figure}[!hp]
\centering
\includegraphics[width=.95\linewidth]{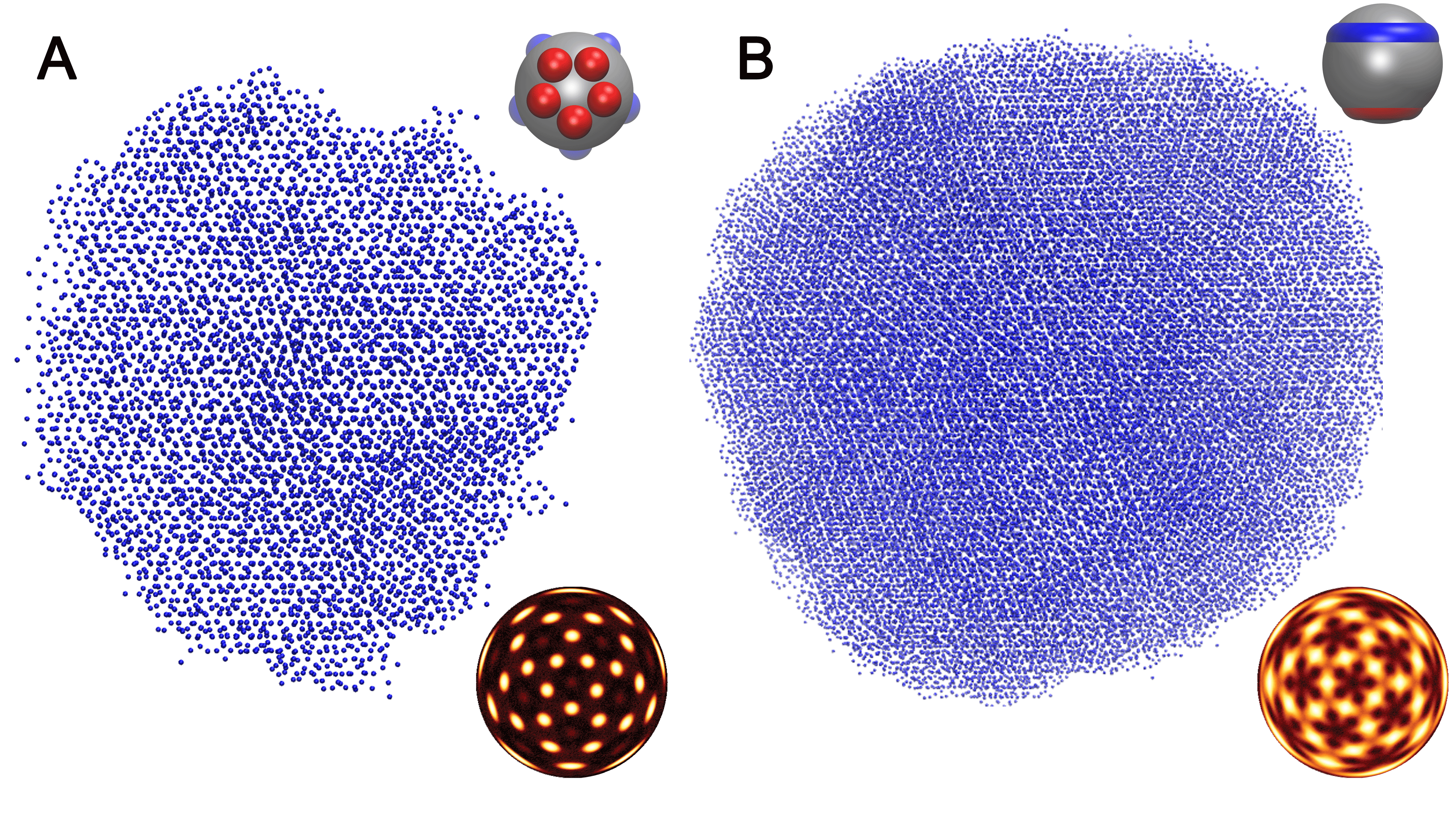}
\caption{Snapshots of ($A$) a 17\,000 particle FCI QC cluster assembled from a gas of {10P} particles and ($B$) a 83\,000 particle FCI QC cluster assembled from a gas of {2R} particles, shown in the top right corners. Particles in the clusters are displayed as small spheres. In these representations the global five-fold order is apparent as well as the presence of phason disorder 
but the absence of dislocations (see also Fig.\ \ref{fig:jags}). As can be seen in the BOOD, bonds are again aligned along the two-fold axes, although they are less constrained in the {2R} system. The diffraction patterns for these IQC  are shown in Supplementary Fig.\ \ref{fig:7P_vs_10P}.
}
\label{fig:FCI_10P}
\end{figure}

\begin{table}[!hp]\centering
\caption{\label{tbl:phason_strain} \textnormal{Phason strain measured as the slope of the distance in perpendicular space between two lifted lattice points as a function of their separation in parallel space (Figs.\ 3J, S5D, S10F and S11I). Fits were measured considering only
distances where the slope of the graph was approximately constant, namely between 40-60\,$\sigma$ in parallel space, except for the {10P} and {2R} models, for which parallel distances between 30-50\,$\sigma$ were considered. 
Phason strain in the {7P} model with $\epsilon_{BB}/\epsilon_{AA}=$1, is about 100 times lower than in the 3/2 approximant and approximately of the same order as that in the annealed ideal FCI QC with the same model. We can conclude that this system has zero phason strain within the accuracy of our calculations. The last column shows the percentage of lifted particles that fall outside the OD, which is set to a triacontrahedron with an inscribed sphere having a radius equal to 2$\sigma$.}}

\begin{tabular}{lcccc}
\hline
System & &Phason strain  & & Points outside OD  \\
 \hline
{7P}, $\sigma=0.30$, $\epsilon_{BB}=$1.0, NT & & 0.0002 & & 8\% \\
{7P}, $\sigma=0.30$, $\epsilon_{BB}=$1.2, NT & & 0.002  & & 6\%\\
{7P}, $\sigma=0.30$, $\epsilon_{BB}=$1.4, NT & & 0.001  & & 8\%\\
{7P}, $\sigma=0.30$, $\epsilon_{BB}=$1.6, NT & & 0.004 & & 6\%\\
\midrule 
{7P}, $\sigma=0.30$, $\epsilon_{BB}=$1.4, WT  & & 0.005 & &  11\% \\
\midrule
Ideal FCI QC, {7P}, $\sigma=0.30$, $\epsilon_{BB}=$1.0, NT & & 0.0001 & & 0\% \\
Ideal 3/2 AC, {7P}, $\sigma=0.30$, $\epsilon_{BB}=$1.0 NT &  & 0.02 & & 0.004\%\\
\midrule
{10P} $\sigma=0.25$, $\epsilon_{BB}=1$, NT & & 0.004   & & 5\% \\
{10P} $\sigma=0.30$, $\epsilon_{BB}=1$, WT & & 0.002  & &  7\% \\
\midrule
{2R} $\sigma=0.15$, $\epsilon_{BB}=1$, NT & & 0.005 & & 13\%\\
\hline
\end{tabular}
\end{table}

\subsection{Structure of the fluid before nucleation}

\begin{figure}[!hp]
\centering
\includegraphics[width=17cm]{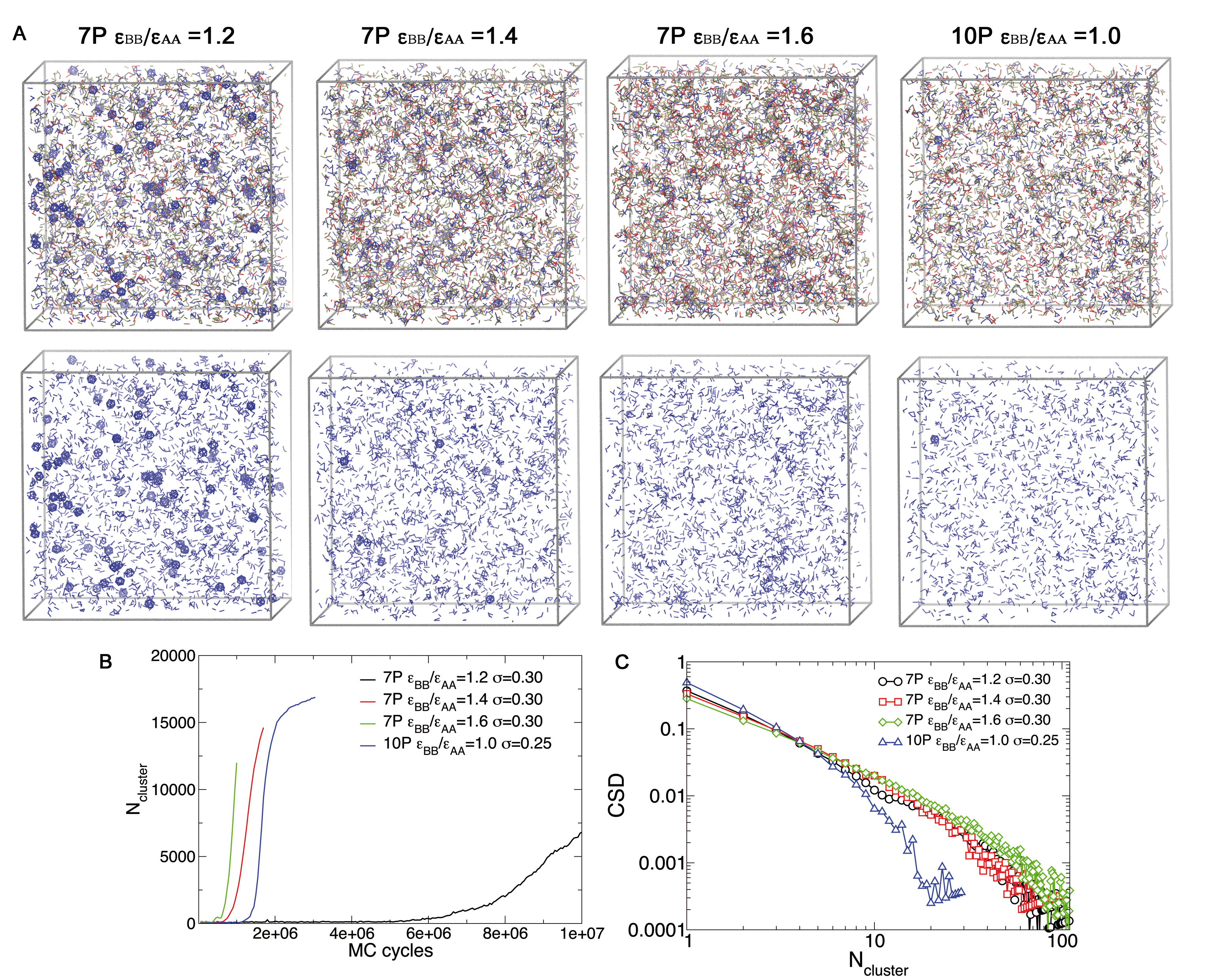}
\caption{\label{fig:fluid_before_nucl} ($A$) Representative snapshots of the fluid  ($\rho^*=0.1$) before nucleation of the IQC in the {7P} and {10P} systems. In the top row, blue lines represent AA (intra-icosahedral) bonds, red lines BB (inter-icosahedral) bonds and tan lines AB bonds. In the bottom row only the AA bonds are shown to aid visualization. The number of icosahedra in the {7P} FCI model decreases as $\epsilon_{BB}/\epsilon_{AA}$ increases. Some icosahedra are visible in the {10P} system, in a proportion that is intermediate between those of the {7P} system with $\epsilon_{BB}/\epsilon_{AA}$=1.4 and $\epsilon_{BB}/\epsilon_{AA}$=1.6. ($B$) Growth rate of the FCI QC cluster from the gas phase. ($C$) Normalized cluster size distribution (CSD) in the fluid before the nucleation of the IQC. The CDS gives the the fraction of particles belonging to a cluster of size $N_\mathrm{cluster}$. In the {10P} system large clusters (above 30 particles) are less frequent than in the {7P} system (for which clusters of up to 100 particles are observed for the three values of $\epsilon_{BB}/\epsilon_{AA}$). For all the model systems, data corresponds to the highest temperature at which nucleation was observed on the time scale of our simulations. }
\end{figure}

With the aim of gaining insight into the mechanism of assembly, we analysed the structure of the fluid before nucleation in some of the model systems. Representative configurations of the fluid and the cluster size distribution (CSD) are presented in Fig.\ \ref{fig:fluid_before_nucl}. As expected, the CSD shows that larger clusters become more frequent as $\epsilon_{BB}/\epsilon_{AA}$ increases. Inspection of the configurations reveals that these large clusters are disordered, and that icosahedra become more frequent as $\epsilon_{BB}/\epsilon_{AA}$ decreases. The presence of icosahedra in the gas phase significantly slows down the growth rate of the cluster (Fig.\ \ref{fig:fluid_before_nucl}B). The {7P} FCI model with $\epsilon_{BB}/\epsilon_{AA}=1.2$ is the model with lowest growth rate. It is plausible that this slow growth provides the required time for defects to anneal out, which may appear when particles in the gas phase adhere to the growing cluster. Indeed, the assembled IQC with this model is the one structurally most similar to the ideal FCI QC and with lowest phason strain (Table\ \ref{tbl:phason_strain}).

\subsection{Estimation of the transition from a gas of particles to a gas of icosahedra}

The transition from a gas of monomers to a gas of icosahedra was investigated by conducting simulations of particles with only five patches of type $A$, i.e.\ the ones involved in intra-icosahedral bonds, thus, allowing us to study this transition unhindered by further assembly of the icosahedra into higher-order structures mediated by the
B patches.
This makes the particles equivalent to those previously studied in Refs.\ \citenum{Wilber07,Wilber09}.
The patch width was set to $\sigma_{ang}=0.3$ radians, as in the {7P} FCI model. Monte Carlo simulations were performed in a cubic simulation box with 20\,000 particles at a density of 0.1\,$\sigma_{LJ}^{-3}$ and at three different temperatures, starting from a random configuration of a gas of monomers. Simulations were run for about 20--30 million MC cycles until equilibrium was reached. The proportion of clusters of sizes between 1 and 12 was monitored for 10 additional million MC cycles. To double-check convergence, separate simulations were run at $T^*=0.158$ starting from a gas of monomers and a gas of mostly icosahedra; the results were effectively identical.

\begin{figure}[!hp]
\centering
\includegraphics[width=16cm]{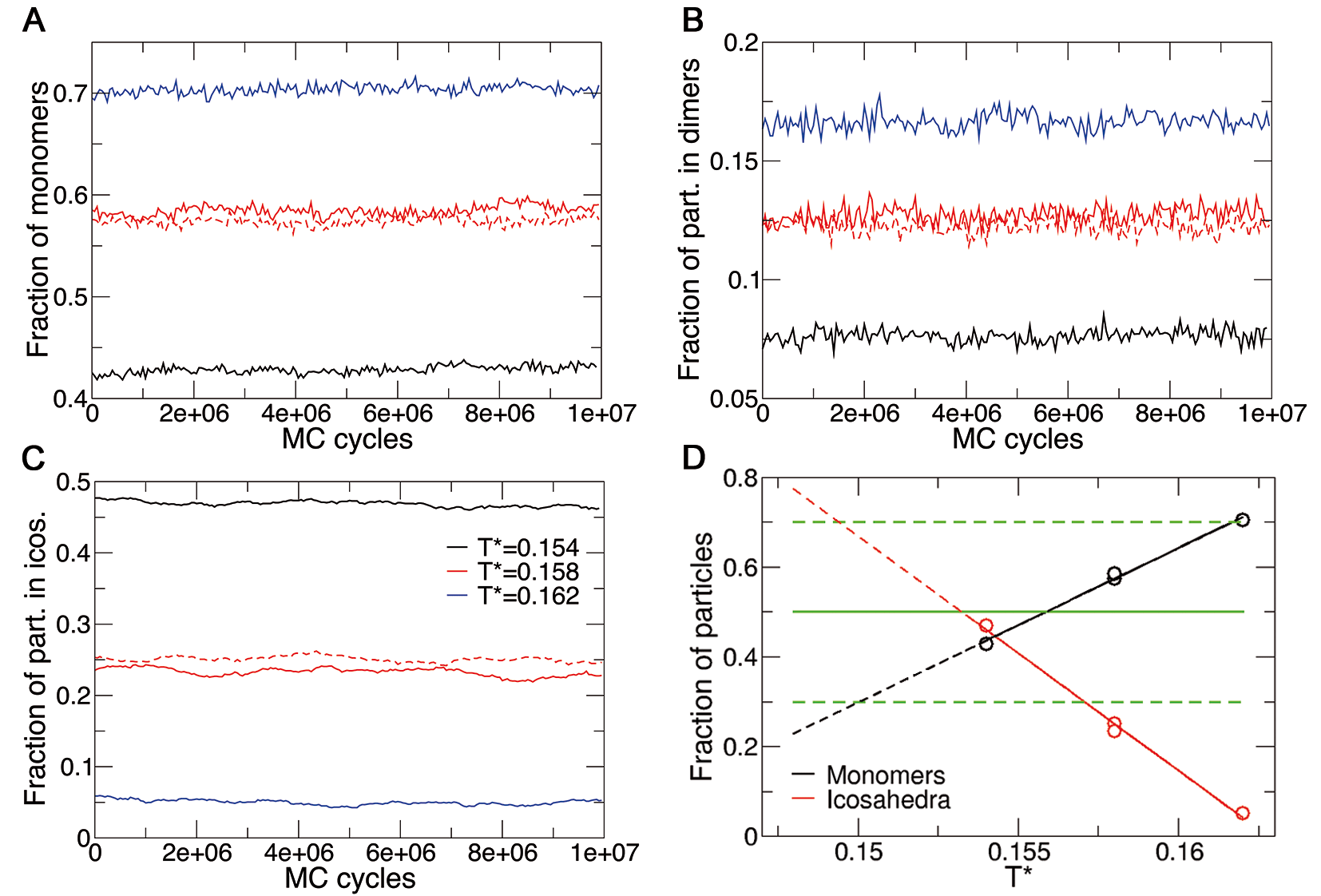}
\caption{\label{fig:icos} (A)-(C) Evolution of the
fraction of particles belonging to (A) monomers, (B) dimers and (C) icosahedra at $T^*$=0.154, 0.158 and 0.162. At $T^*=$0.158 two trajectories are shown: the straight line corresponds to a simulation starting from a random disordered configuration and the dashed line was obtained starting from an equilibrated configuration at $T^*=0.154$, in which a large proportion of particles are forming icosahedra. (D) Fraction of
particles belonging to monomers and to icosahedra as a function of temperature. The lines are linear fits. The horizontal green solid and dashed lines mark the 0.5 and 0.5$\pm$0.2 probabilities. The transition temperature from a gas of particles to a gas of icosahedra was taken as that at which a 0.5 fraction of particles are forming icosahedra.  }
\end{figure}

The centre of the transition from a gas of monomers to a gas of icosahedral clusters is estimated to occur at $T^*=0.153$ and is where 50\% of particles are in icosahedra. 
This result is lower than the value of $T^*$=0.156 
that was obtained by umbrella sampling Monte Carlo simulations of 5-patch particles in a system containing 12 particles \cite{Wilber07}; however, that study neglected finite-size effects \cite{Ouldridge10b}, whereas for the large systems used here these effects are negligible.
The temperatures at which 70\% ($T^*=0.149$) and 30\% ($T^*=0.157$) of particles assemble into icosahedra provides an indicator of the width of the transition (see Fig\ \ref{fig:icos}).

\subsection{Nucleation and growth mechanisms}

Although conducting a rigorous study on the nucleation and growth mechanisms of the FCI QC is beyond the scope of this work, some insights can already be gained from a visual inspection of the trajectories. Two different nucleation and growth mechanisms were observed at ($\epsilon_{BB}/\epsilon_{AA}=1.2$) and above ($\epsilon_{BB}/\epsilon_{AA}=1.4$) the transition from a gas of particles to a gas of icosahedra of the fluid. These are shown in the Supplementary Movies M1 and M2. The movies correspond to the trajectories shown in Fig.\ S8. As can be seen, the growth rate is significantly slower for the model with $\epsilon_{BB}/\epsilon_{AA}=1.2$ than for that with $\epsilon_{BB}/\epsilon_{AA}=1.4$ .  The movies show the evolution of the largest cluster from its birth until it reaches a fairly large size. For clarity, only bonds (not particles) are shown. These are depicted in different colours depending on the assembled state of each particle: particles in the largest solid cluster identified with an order parameter (similar to that used by Keys and Glotzer\cite{Keys07}, but with $q_{12}(i)q_{12}(j)$  as this exhibits a distinctive signal for the IQC) are shown in blue, those bonded to the largest cluster but not identified as quasicrystalline are shown in purple, and those pertaining to clusters of bonded particles with sizes larger than 10 particles are shown in green. 

As can be seen in the Supplementary Movie M1, for the system with $\epsilon_{BB}/\epsilon_{AA}=1.2$, many icosahedra are present in the fluid phase before nucleation. Some of these icosahedra bind to each other, initially not necessarily with the right orientation or arrangement of B patches on their surface, but as the simulation progresses, they are able to reorient and rearrange and thus promote growth. This is a rather slow process, and several crystallites compete with each other. Eventually, one of the crystallites outweighs the others and continues to grow by attachment of further icosahedra from the fluid phase. There are also a significant number of particles bonded to the surface of the solid cluster which are not part of complete icosahedra and are likely also contributing to cluster growth. Overall, we can conclude that in this case, cluster growth is primarily driven by the addition of icosahedra from the fluid phase. By contrast at $\epsilon_{BB}/\epsilon_{AA}=1.0$, the icosahedra are insufficiently dynamic to allow growth of a large IQC. 

The mechanism in the system with $\epsilon_{BB}/\epsilon_{AA}=1.4$ is notably different (Movie S2). In this case, there are far fewer icosahedra in the fluid phase before nucleation. Particles still tend to form relatively large aggregates, but these are now more disordered, with only a few icosahedra present. In this example, growth begins from a dimer of icosahedra, which, over time, grows primarily through single-particle addition. However, the addition of a few icosahedra is also observed during cluster growth. Single particles can align more quickly and coherently with the orientational symmetry of the formed cluster than the icosahedra, which is why cluster growth is significantly faster in this system than in the previous one.

\subsection{Comparison of the {7P}, {10P} and {2R} FCI models}

\begin{figure}[!hp]
\centering
\includegraphics[width=13.1cm]{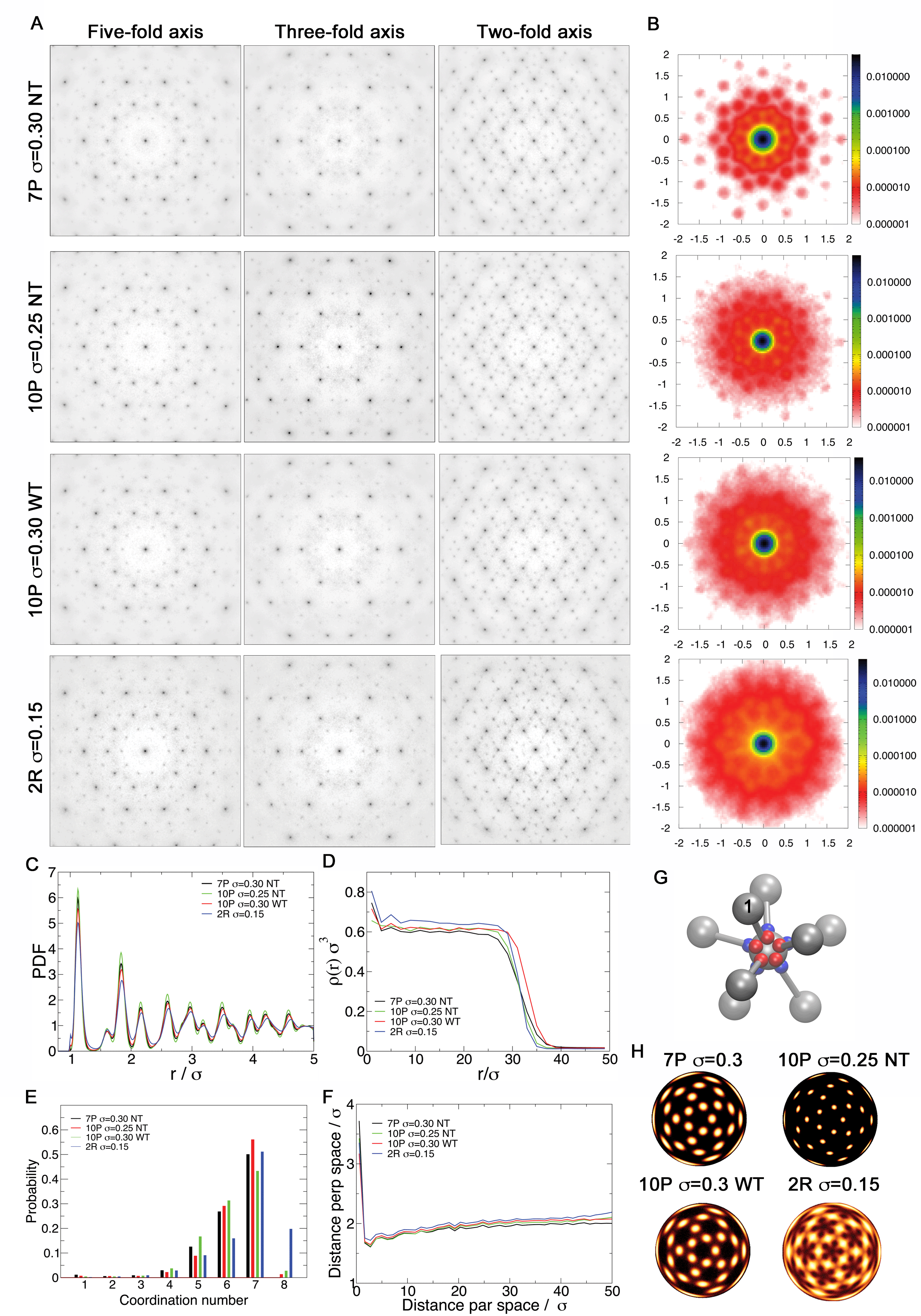}
\caption{\label{fig:7P_vs_10P} Comparison of the IQC assembled with the {7P}, {10P} and {2R} models. ($A$) Projections of the diffraction pattern along 5-fold, 3-fold and 2-fold axes. ($B$) Van Hove correlation function. ($C$) Pair distribution function. ($D$) Radial density. ($E$) Coordination number distribution. ($F$) Phason strain. ($G$) An example particle bonded to 8 neighbours extracted from the {10P} $\sigma=0.25$ NT system. The Particle with the label "1" is able to squeeze inbetween two particles bonded to the $B$ patches to form an extra weak bond, in this particular example the energy of the bond is $-0.40$\,$\epsilon$. We speculate that these defects help to make the {10P} QC more `dynamic', as shown by the Van Hove correlation plots.
($H$) BOOD diagrams. Results for the {7P} FCI model correspond to $\epsilon_{BB}/\epsilon_{AA}=1.2$.}
\end{figure}

The structural properties of the assembled IQCs with the {7P}, {10P} and {2R} FCI models are compared in Fig.\ \ref{fig:7P_vs_10P}. The {7P} and {10P} models form structurally very similar IQCs. The brightest peaks in the diffraction pattern are shared by these two model systems. The PDFs are also very similar. The only appreciable difference is that peaks in the {10P} FCI system with $\sigma_{ang}=0.25$ radians are more defined, which is attributed to the narrower patch width. The density of the {10P} IQCs is slightly larger than that of the {7P} IQC, which can be attributed to the appreciable number of particles with eight nearest neighbours in the {10P} system (Fig.\ \ref{fig:7P_vs_10P}E). If all neighbours are located at the minimum energy position, only seven nearest neighbours are possible due to geometric constrains. However, up to eight bonds are possible if the neighbour particles are slightly displaced from the energy minimum positions (Fig.\ \ref{fig:7P_vs_10P}G). This comes with an energy penalty and, indeed, when coordination numbers are calculated using an energy criterion {7P} and {10P} IQCs exhibit similar average coordination numbers (Table\ \ref{tbl:bond_type}), but applying a distance criterion $\langle CN \rangle$ is about 1\% lower in the {10P} IQCs.

 When the two sets of patches are replaced by two ring patches, the diffraction pattern again exhibits the main features of the {7P} and {10P}, although there might be some changes in the peaks' relative intensities. Some small differences are also visible in the PDF. The peaks become somewhat lower and broader in the {2R} QC. The radial density is about 2-3\% higher in this case, which might be related to a significant number of particles with eight neighbours (about 20\% have coordination eight, Fig.\ \ref{fig:7P_vs_10P} $E$). The BOOD is also very different from those in the {7P} and {10P} QCs. The bright spots become much broader, indicating that bonds are more loosely constrained to the 30 directions of the $C_2$ symmetry axes of the $I_h$ point group than in the {7P} and {10P} QCs. The phason strain is somewhat larger than to those of the {7P} and {10P} models (Table\ \ref{tbl:phason_strain}), but the number of unassigned particles due to conflicting assignments is significantly larger than in the remaining particle designs (about 10\% cannot be lifted, as compared to less than 1\% in the {7P} and {10P} systems). This is likely due to the greater disorder in this system.

\subsection{Simulations of the ideal FCI and 3/2 approximant with the {7P} model}

\begin{figure}[!hp]
\centering
\includegraphics[width=14.3cm]{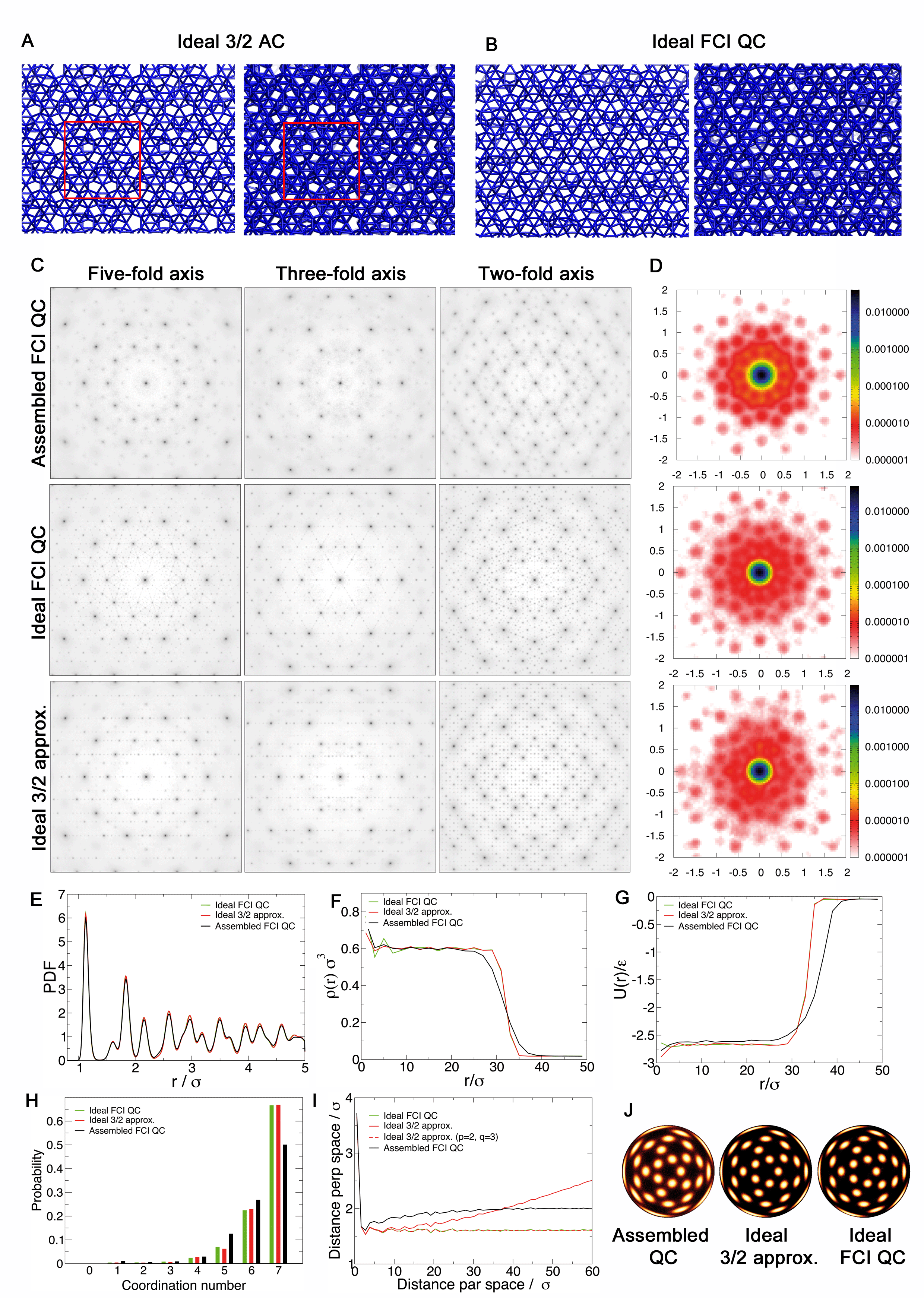}
\caption{\label{fig:annealed_samples} Annealed ideal FCI QC and a 3/2 approximant versus the assembled IQCs. The three systems were simulated using the {7P} design with $\epsilon_{BB}/\epsilon_{AA}=1$ at $T^*=0.154$. Structure of the pristine and tempered ($A$) 3/2 AC and ($B$) ideal FCI QC. ($C$) Projections of the diffraction pattern along 5-fold, 3-fold and 2-fold axes. ($D$) Van Hove correlation function. ($E$) Pair distribution function. ($F$) Radial density. ($G$) Radial energy. ($H$) Coordination number distribution. ($I$) Phason strain. ($J$) BOOD diagrams. }
\end{figure}

A complete structural and dynamic characterization of the annealed ideal FCI QC and the 3/2 approximant simulated with the {7P} particle design with all patches having equal interaction strength $\epsilon_{BB}/\epsilon_{AA}=1$ is provided in Fig.\ \ref{fig:annealed_samples}.

\subsection{Particle mobility during the simulations: The evolution of the Van Hove correlation function}

\begin{figure}[!hp]
\centering
\includegraphics[width=17cm]{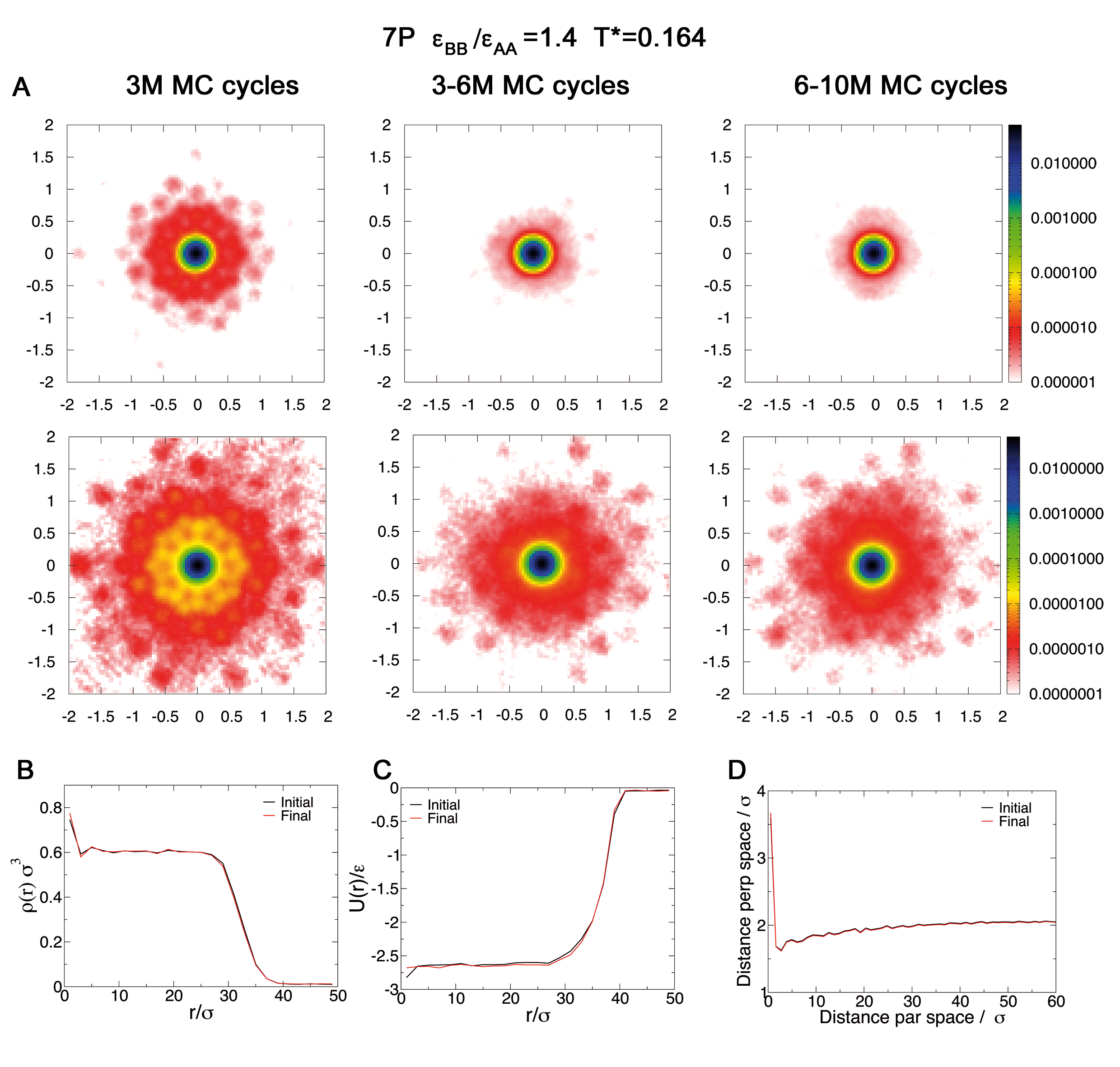}
\caption{\label{fig:VanHove} ($A$) Evolution of the Van Hove correlation function along a 10 million MC cycles trajectory of the {7P} FCI model with $\epsilon_{BB}/\epsilon_{AA}=1.4$ at $T^*$=0.164. The Van Hove correlation function was evaluated after 1 million MC cycles, averaging over 3--4 million MC cycles. Only particles within a radial distance of 27\,$\sigma$ of the centre of the condensed cluster were included in the evaluation of the Van Hove function to avoid surface effects. The two rows show the same data, but in the bottom row probabilities from 10$^{-7}$ are shown (as compared to 10$^{-6}$ in the upper row) to reveal more features of the diagrams. Particle mobility is significantly larger in the initial stage of the simulation (3 M MC cycles), after which it reaches a steady state. ($B$) The radial density of the cluster is virtually the same at the beginning and at the end of the 10 million MC cycles simulation. ($C$) The radial energy at the beginning and at the end is also very similar, with a very mild decrease at radii above 20\,$\sigma$, close to the cluster surface. ($D$) The phason strain remains also almost invariant along the simulation. 
}
\end{figure}

The mobility of the particles along the simulations was quantified by evaluating the Van Hove correlation function, measured after 1 million MC cycles in the {7P} FCI system with $\epsilon_{BB}/\epsilon_{AA}=1.4$. Evaluation of the correlation at different stages of a long annealing simulation reveals that particle mobility is higher at the early stages of the simulation, but then reaches a steady state of lower mobility (Fig\ \ref{fig:VanHove});  this change is probably due to the annealing out of defects.
Note however that probabilities of particle hops after 1 million MC cycles are quite small and do not lead to significant relaxation of the structure, even after the 10 million long simulation. Both the radial density and energy and the phason strain are almost indistinguishable at the beginning and at the end of the simulation (Fig\ \ref{fig:VanHove}).

\begin{figure}[!hp]
\centering
\includegraphics[width=9cm]{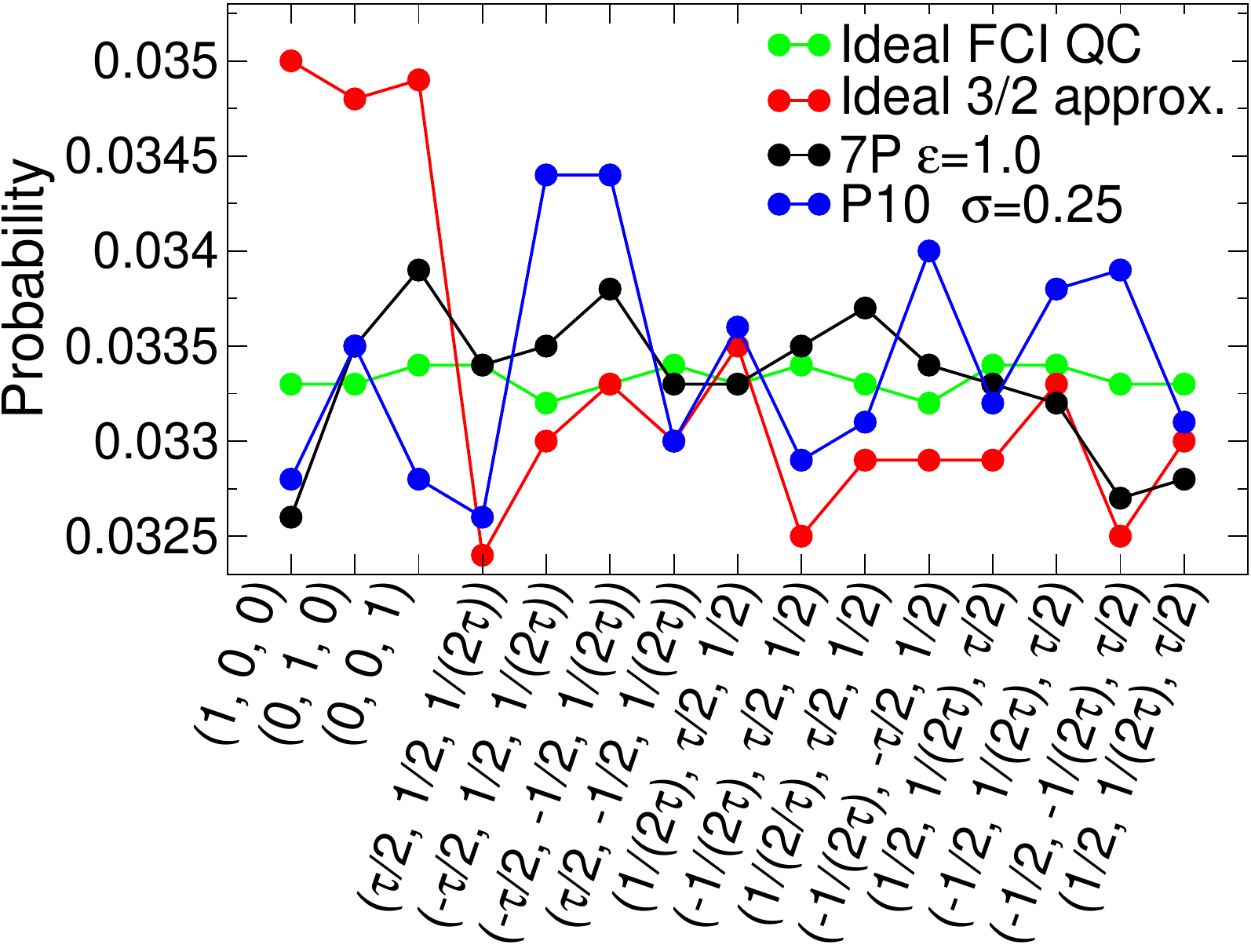}
\caption{Integral of the bright spots in the BOODs along each of the 15 2-fold axes. Unsurprisingly, the BOOD for the annealed ideal FCI QC has the most uniform distribution. For the approximant there is a clear higher probability of bonds being oriented along the $x$, $y$ and
$z$ axes that breaks the $I_h$ symmetry.}
\end{figure}

\begin{figure}[!hp]
\centering
\includegraphics[width=16cm]{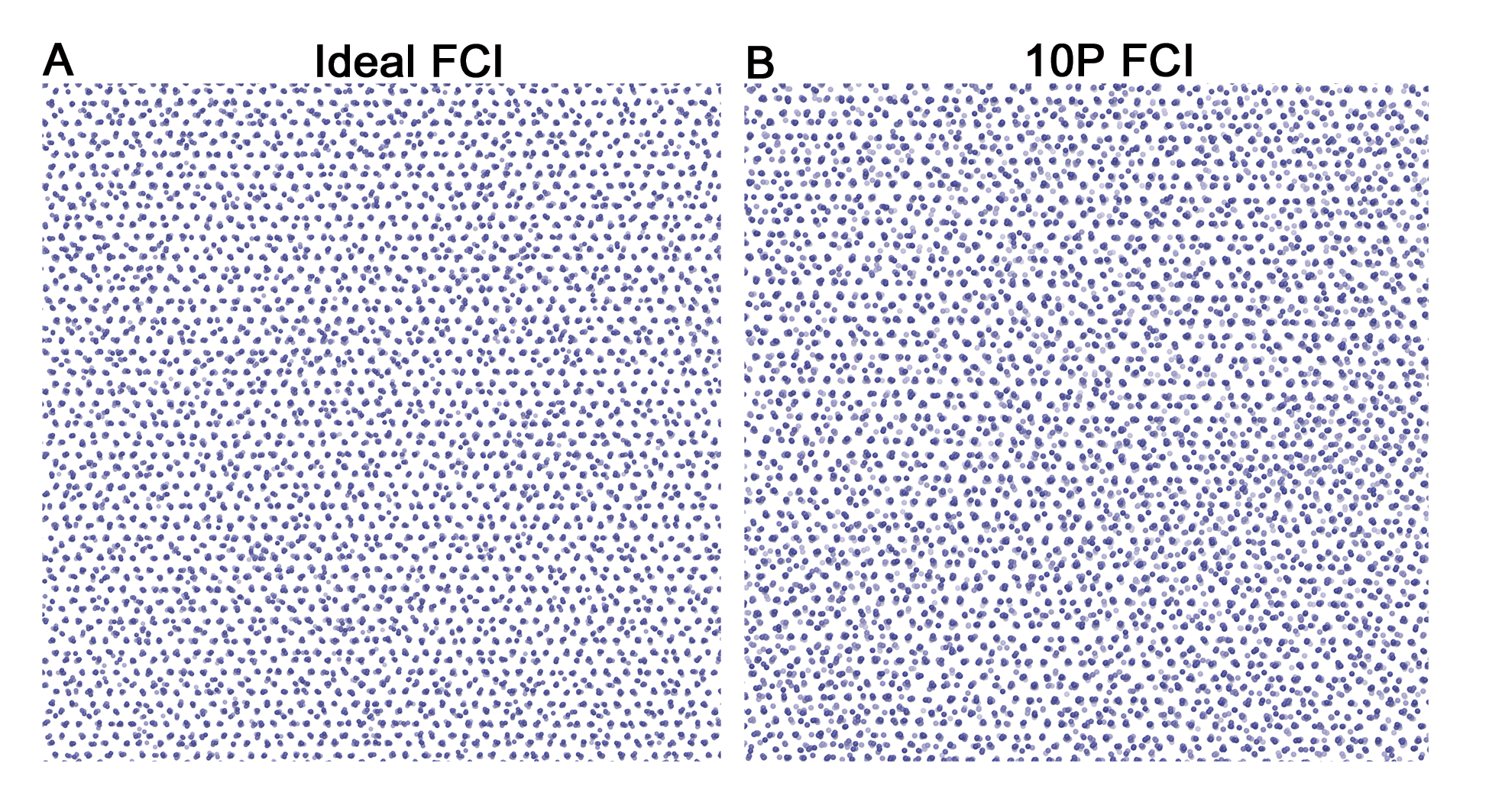}
\caption{\label{fig:jags}Comparison of the structures of the (A) tempered ideal FCI QC and (B) assembled {10P} FCI QC. The {10P} FCI QC possesses significantly greater phason disorder. This can be observed by inspecting the ``lines'' (of particles or gaps between particles) that run through the structures at 0, $\pm 36^\circ$, $\pm 72^\circ$, $\pm 108^\circ$ and $\pm 154^\circ$  to the horizontal (these are most easily seen viewed from a glancing angle). In the ideal QC each of these lines mostly run completely through the structure, 
but in the assembled they are often interrupted part way through by a ``shift'' in the structure (this shift is often termed a jag \cite{Socolar86}). 
However, that the lines in a given direction are almost perfectly parallel provides evidence of the absence of dislocations (further confirmed by the lifting analysis).}
\end{figure}

\subsection{Further results of lifting}

\begin{figure}[!h]
\centering
\includegraphics[width=12.7cm]{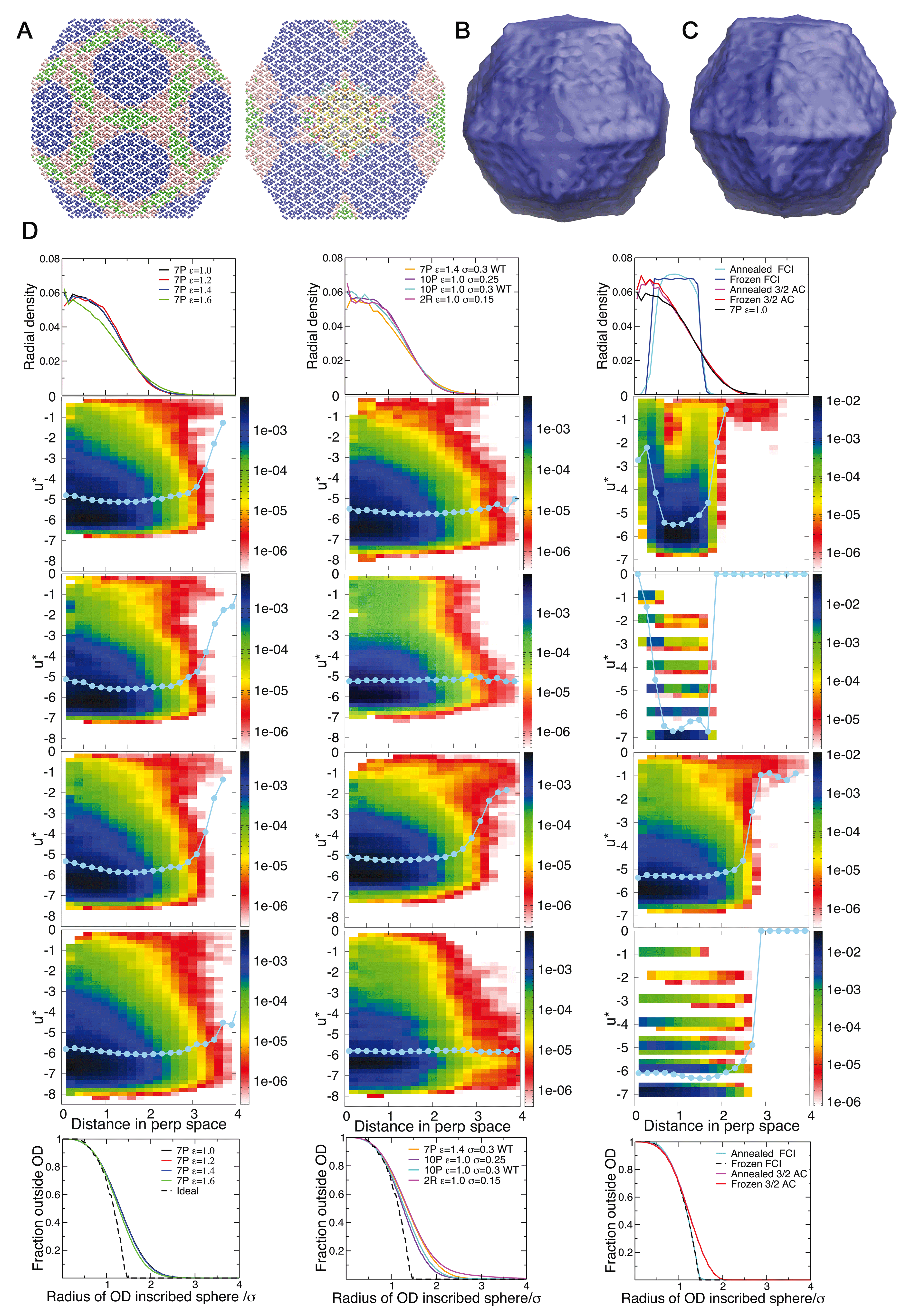}
\caption{\label{fig:lifting} (A) Occupation domain of the ideal FCI QC. Each point is colored according to its local environment in real space (using the same color scheme as that in Fig.\ 1 of the main text). In the right-hand image the occupation domain is bisected to aid visualization. Isosurfaces of the occupation domain (B) in the ideal FCI QC and (C) in a tempered simulation starting from the ideal FCI QC. (D) Density of points in perpendicular space as a function of distance in perp space (top panel) and as a function of distance in perp space and particle energy ($u^*$). The energy-distance plots are shown from top to bottom in the same order as the legend of the radial density plots (top panel). Bottom panels show the fraction of points outside a `scaled' occupation domain as a function of the radius of the inscribed sphere of the scaled dodecahedral occupation domain.}
\end{figure}

\begin{figure}[!h]
\centering
\includegraphics[width=14.0cm]{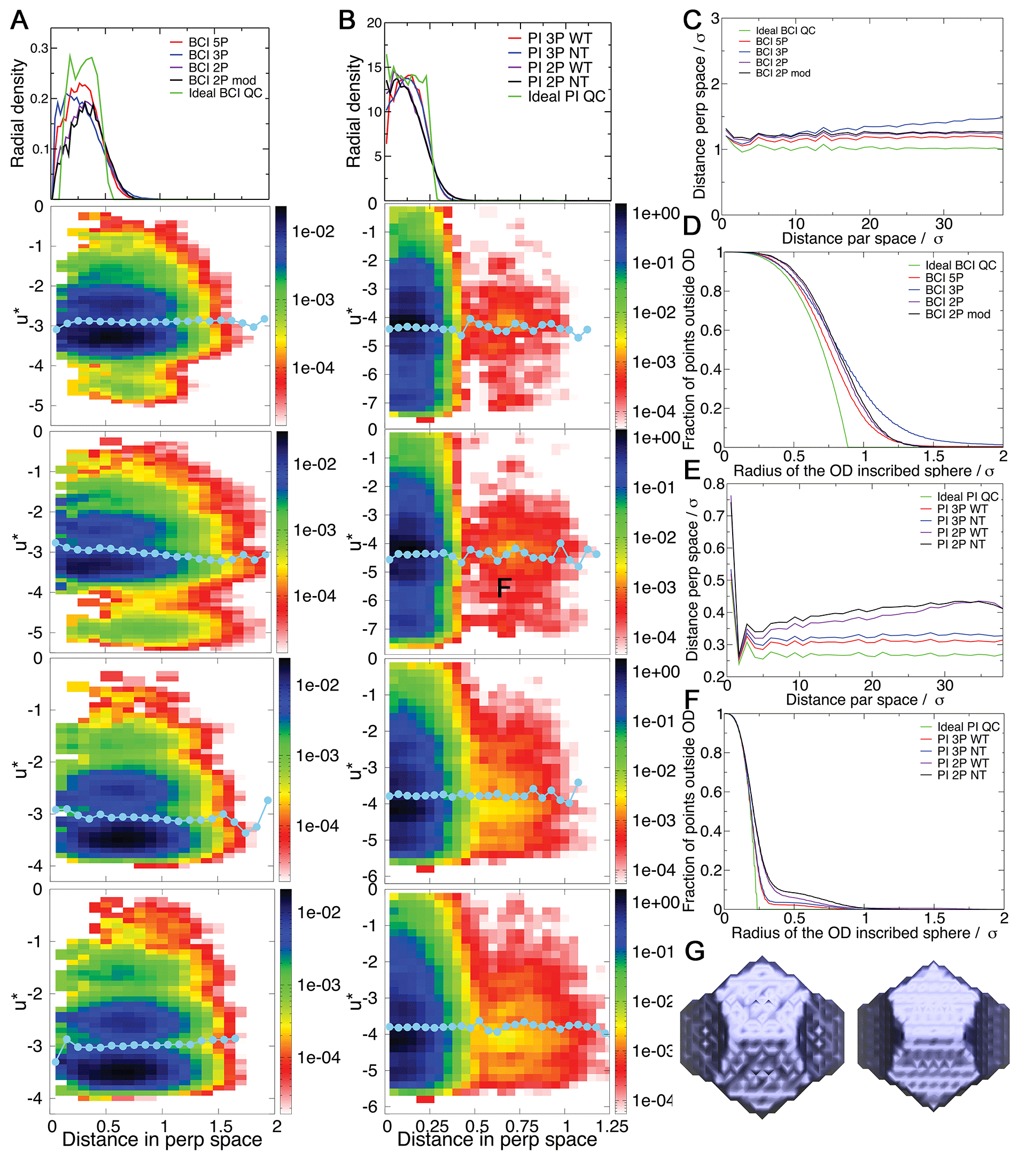}
\caption{\label{fig:lifting-BCI-PI} Lifting results for the BCI and PI QC reported in our previous work\cite{Noya21}. (A) and (B): Density of points in perpendicular space as a function of distance in perp space (top panel), and as a function of distance in perp space and particle energy ($u^*$). The energy-distance plots are shown from top to bottom in the same order as the legend of the radial density plots (top panel). (C) Phason strain for the BCI QC. (D) Fraction of points outside a `scaled' occupation domain as a function of the radius of the inscribed sphere of the scaled dodecahedral occupation domain for the BCI QC model systems. This quantity was calculated by scaling the dodecahedral occupation domain by a given factor and measuring the number of atomic positions that fall outside this occupation domain. (E) Phason strain for the PI QC systems. (F) Fraction of points outside a `scaled' occupation domain as a function of the radius of the inscribed sphere of the scaled dodecahedral occupation domain for the PI QC systems. (G) Shape of the occupation domain for the ideal BCI (left) and PI (right) QC. In both cases it has dodecahedral shape. Figures do not respect the relative size between both OD.}
\end{figure}

Additional information about the occupation domain (OD) of the ideal FCI and 3/2 approximants as well as that of the assembled QCs obtained from the lifting procedure is provided in Fig.\ \ref{fig:lifting}. For completeness, we have also applied the lifting analysis to the primitive (PI QC) and body-centred (BCI QC) icosahedral quasicyrstals obtained in our previous work\cite{Noya21} (Fig.\ \ref{fig:lifting-BCI-PI}). The results are somewhat similar to those of the FCI QCs. Most of the assembled QCs exhibit very low phason strain, except for the BCI 3P and PI 2P systems. As in the FCI QCs, the energy of the particles is again rather independent of their distance from the center of the occupation domain. Curiously,  two clearly distinct regimes can be observed in the plots of percentage of points outside the OD as a function of the radius of the inscribed sphere for the assembled PI QCs (Fig.\ \ref{fig:lifting-BCI-PI}F): there is an initial sharp drop coinciding with that of the ideal PI QC, followed by a much slower decay at larger distances from the center of the OD. 

\section{Potential experimental realization}

\subsection{DNA nanotechnology}

The two most successful approaches to produce 3-dimensional periodic crystals made from DNA origami have been to use DNA origami polyhedra with single-stranded ``sticky ends'' at their vertices \cite{Tian20,Liu24} or multi-arm DNA origami particles \cite{Posnjak24}. In the former approach each edge of the polyhedron is a multi-helix bundle and the single strands that mediate the interactions have a linker section and a complementary binding region. Due to the flexible linkers, the strands mediate little further angular constraints on the vertex-vertex interactions. As an example of the latter approach, in Ref.\ \citenum{Posnjak24} 4-arm DNA origami particles were able to assemble into a diamond lattice. Each arm consisted of a 24-helix bundle. At the centre of the design each arm splits into three 8-helix bundles that merge with the neighbouring arms. ``Insertions'' and ``deletions'' were incorporated into the arms at these points to facilitate the required bending \cite{Dietz09}. The symmetry of the design ensures the tetrahedrality of the particles. Six of the helices at the ends of the arms had short single-stranded extensions that mediated the interarm bonding; the pattern of extensions was designed to provide torsional control, in particular to ensure the ``staggered'' bonding geometry required for the diamond lattice. 

Probably the most natural way to generate an equivalent of the patchy particles in the current paper that has the correct patch geometry is to use a single DNA origami particle with approximate 
(DNA origami have $C_1$ point group symmetry when all microscopic details are considered) icosahedral symmetry. In particular, the ``wireframe'' origami approach has been used to generate single DNA origami icosahedra with two helices along edge \cite{Veneziano16} and more recently with each edge being a 4-helix bundle \cite{Zhang22,Liu24}. 
The extra rigidity of the 4-helix bundle icosahedron probably makes it more suitable for higher-order assembly and it was one of the polyhedra used in Ref.\ \citenum{Liu24} to generate DNA origami crystals. 

The appropriate 7-edges of such a DNA icosahedron could be functionalized with single-stranded sticky ends to generate interactions that mimic the patchy particles considered here. 
It would make sense to use multiple sticky ends per edge. These would not all need to be at the centre of the edge, but could also be symmetrically  positioned about the centre, as long as the sticky ends were specific enough to only interact with strands that are placed at the same distance from the centre. In this case, this would naturally generate a preference for edge-to-edge bonding, i.e.\ there would be some torsional-like constraint, the degree of which would depend on length of linkers. Note that we have shown that the patchy particles can assemble into an icosahedral QC both with and without a torsional component to the interactions and the edge-to-edge character of the interactions would give a preferred torsional angle that matches that required (Table \ref{tbl:7p_fci_model}). In contrast to this torsional constraint, the rocking motion about an axis parallel to the edges would be relatively free. 
Note also that this design could allow AB inter-patch interactions as well as AA and BB interactions, whilst also allowing the relative strengths of the BB and AA interactions to be fine tuned (for example, by having different numbers of sticky ends on the A and B edges, or by varying the length of the complementary sections).
Another potential hurdle is the self-complementarity of the interactions that are necessary in a one-component system, as the flexible linkers mean that intra-origami sticky-end binding might be more favourable than inter-origami binding. However, an approach to overcome this issue has been recently demonstrated that allows the assembly of crystals in one-component systems of DNA octahedra. \cite{Zhou24c}

An alternative approach might be to use 7-arm origami particles similar to the 4-arm particles in Ref.\ \citenum{Posnjak24}. However, it would be more difficult to get the correct geometry than for particles explicitly based on an icosahedral geometry. In particular, the inter-arm angles would need to be tuned via ``insertions'' and ``deletions'' \cite{Dietz09} in order to best match the desired geometry. Modelling tools, such as oxDNA \cite{Snodin15,Snodin19}, could play a significant role in this fine-tuning of the design. Postive features are that these particles are likely to be more rigid and have interactions that have a directionality that is more similar to our patchy particles. The interactions would also be torsionally-specific.

 The above discussion is focussed on the 7P system. Although DNA origami analogues of the 10P particles could also be produced in a similar manner to suggested above, the requirement that the excluded volume limits the number of interactions with the B patches could be more problematic to implement. In particular, the icosahedra might not be sufficiently rigid to prevent this and the multi-arm particles would have a excluded volume that is very different from the spherical particles we consider here.


\subsection{Protein design}

Another possibility to realize the quasicrystals in this paper would be through the tools of protein design, which have advanced considerably in recent years, especially as a result of approaches that leverage the power of machine learning.\cite{Winnifrith24}

Although the propensity of many proteins to crystallize under the appropriate conditions has been vital to the field of protein structure determination, this is usually more an unintended consequence of their well-defined structure. However, although comparatively rare, there are a significant number of examples of proteins that have evolved to crystallize for some functional purpose.\cite{Doye06b,Schonherr18} Similarly, proteins can be designed to form two- and three-dimensional periodic crystals \cite{Yeates17,Zhu21b}.

A recent example of the state-of-the-art in the design of proteins that can crystallize comes from Ref.\ \cite{Li23}. The approach is to first design proteins that can form high symmetry complexes that match the local symmetry of high symmetry sites in a crystal. Then inter-complex interactions are designed that allow these complexes to further assemble into the desired crystal. One example of this hierarchical approach was the creation of a diamond-like lattice in which a binary 24-protein complex with tetrahedral symmetry further assembles via intercomplex interactions along the four 3-fold axes of the tetrahedral complex.\cite{Li23}

Like with DNA origami, one potential way to generate a particle with the correct ``patch'' geometry might be to start from a complex with icosahedral symmetry, there being a number of such designed examples \cite{Lutz23,Lee24,Huddy24} (as well as of course natural examples, for example, many virus capsids). The simplest ($T=1$) involve 60 proteins.
However, having intercomplex interactions associated with only a subset of the $C_2$ sites in the icosahedral complex would be a more complex task. One would need to first introduce some kind of programmed symmetry breaking where the icosahedron would be made of multiple homologous proteins with recoded interactions to generate the necessary specificity. Analogues of the $C_{5v}$ P10 particle (rather than the $C_s$ P7 particles) would likely be easier to achieve as it would require fewer distinct proteins due to the higher symmetry. Such programmed symmetry breaking has recently been used to generate a designed $T=4$ icosahedra (and other polyhedra) \cite{Lee24}. Furthermore, there has also been recent progress in generating designed multi-component complexes by the use of standardized protein building blocks \cite{Huddy24}. 

These challenges would seem to suggest that the realization of a protein icosahedral quasicrystal equivalent to those introduced in the current paper would be a harder task than one made from DNA origami, but one should not underestimate the rapid progress being made in the field of protein design.

\FloatBarrier






\providecommand{\latin}[1]{#1}
\makeatletter
\providecommand{\doi}
  {\begingroup\let\do\@makeother\dospecials
  \catcode`\{=1 \catcode`\}=2 \doi@aux}
\providecommand{\doi@aux}[1]{\endgroup\texttt{#1}}
\makeatother
\providecommand*\mcitethebibliography{\thebibliography}
\csname @ifundefined\endcsname{endmcitethebibliography}
  {\let\endmcitethebibliography\endthebibliography}{}